\documentclass[12pt]{article}
\usepackage{epsfig,amsfonts,amssymb}
\usepackage{hyperref}
\usepackage{cite}
\input epsf.sty
\usepackage{cite}

\def\ZZZ{{\hbox{ Z\kern-1.6mm Z}}}
\def\RRR{{\hbox{ R\kern-2.4mm R}}}
\def\CCC{{\hbox{ C\kern-2.3mm C}}}
\def\zzz{{\hbox{z\kern-1mm z}}}

\newcommand{\nn}{\nonumber \\}

\newcommand{\qeq}{{\hbox{=\kern-2.3mm ? \kern.5mm }}}
\renewcommand{\qeq}{=}

\newcommand{\eps}{\epsilon}

\newcommand{\NN}{{\cal N}}

\newcommand{\be}{\begin{equation}}
\newcommand{\ee}{\end{equation}}
\newcommand{\ben}{\begin{eqnarray}\displaystyle}
\newcommand{\een}{\end{eqnarray}}

\newcommand{\refb}[1]{(\ref{#1})}

\newcommand{\sectiono}[1]{\section{#1}\setcounter{equation}{0}}

\def\one{{\hbox{ 1\kern-.8mm l}}}
\def\zero{{\hbox{ 0\kern-1.5mm 0}}}

\newcommand{\bea}[1]{\begin{eqnarray}\label{#1} }
\newcommand{\eea}{\end{eqnarray}}

\begin{document}

\baselineskip 24pt

\begin{center}
{\Large \bf  BPS Spectrum, Indices and Wall Crossing
in $\NN=4$ Supersymmetric Yang-Mills Theories}

\end{center}

\vskip .6cm
\medskip

\vspace*{4.0ex}

\baselineskip=18pt

\centerline{\large \rm Ashoke Sen}

\vspace*{4.0ex}

\centerline{\large \it Harish-Chandra Research Institute}
\centerline{\large \it  Chhatnag Road, Jhusi,
Allahabad 211019, India}

\vspace*{1.0ex}
\centerline{\small E-mail:  sen@mri.ernet.in}

\vspace*{5.0ex}

\centerline{\bf Abstract} \bigskip

BPS states in $\NN=4$ supersymmetric $SU(N)$ gauge theories in
four dimensions can be
represented as planar string networks with ends lying on D3-branes. 
We introduce several protected indices which capture information on the
spectrum and various quantum numbers of these
states, give their wall crossing formula and describe how using the
wall crossing formula we can compute all the indices at all points in the
moduli space.

\vfill \eject

\baselineskip=18pt

\tableofcontents

\sectiono{Introduction} \label{s0}

BPS states on the
Coulomb branch of $\NN=4$ supersymmetric Yang-Mills theories 
in four space-time dimensions have provided
us with important tools for understanding various non-perturbative aspects
of the theory\cite{wittenolive,osborn,9402032,segalselby,9712211,9804160}. 
A convenient way to represent these states in $SU(N)$ gauge theories
is to regard the gauge theory as the world-volume theory of $N$ D3-branes.
The BPS states are then described as planar networks of open strings
ending on D3-branes\cite{9712211,9804160}. The half BPS states
correspond to single open strings stretched between a pair of D3-branes
whereas quarter BPS states correspond 
to more general planar string network
constructed by joining many three string 
vertices\cite{9607201,9704170,9710116,9711094,9711130,9812021}. 
Of these the spectrum of half BPS
states is by now completely understood and can be obtained simply by a 
duality transformation of the spectrum of the massive gauge bosons.
However 
despite the simple representation of the quarter BPS states as string
network on D3-branes, and several extensive studies of the 
spectrum of these states
from the study of bound states of multiple 
monopoles\cite{9804174,9907090,9912082,0005275,0609055} and other
techniques\cite{0712.3625,0802.0761}
the full
spectrum of quarter BPS states is still unknown.
The goal of this paper will be to provide a complete answer to this
problem.

We shall address the problem in two steps. The first step will be to introduce
appropriate protected indices which get contribution from BPS states but
not from  a non-BPS state. These indices contain information on the
BPS spectrum which are stable under 
quantum corrections. 
In the second step we shall describe the wall crossing 
formula\cite{9211097,9407087,
9408099,9602082,9605101,9902116,0011017,
0005049,0206072,0304094,0702141,0702146,0702150,
0705.3874,0706.2363,0811.2435,
0910.4315,
1006.2706,0410268,0810.5645,0910.0105,0706.3193,
0807.4723,0812.4219,0807.4556,
0810.4909,0904.1420,0910.2615,0912.2923,0912.2507,1002.0579,
1006.3435,1006.0146,1008.0030,
0906.1767,0912.1346,1011.1258,1003.1570,1009.1775,
1010.6002,1102.1729,1103.0261,
1103.1887,1107.0723,1109.4861,1109.4941,1110.0466,1110.1619,
1111.6979,1112.2174,1112.2515}  for these
indices.\footnote{Application of wall crossing formula to special configurations
in $\NN=4$ supersymmetric gauge theories can be found in
\cite{0802.0761}.}
Combining the wall crossing formula with the
observation that for collinear configuration of D3-branes there are no
BPS string network except the half BPS states, we can derive the
formula for the index at a generic point in the moduli space by identifying
the walls of marginal stability which need to be crossed as we deform the
moduli from the collinear configuration to the configuration of interest. 
This is
a purely kinematic problem and can in fact be solved diagrammatically by
following the deformation of the string network configuration as we deform
the locations of the D3-branes.

We now summarize our main results and the organization
of the paper. After reviewing the string 
network representation of BPS states in \S\ref{s1.5}
we introduce in \S\ref{s1}
three different indices for the BPS states in $\NN=4$
supersymmetric gauge theories. The first one is the standard sixth
helicity trace index $B_6$\cite{9611205,9708062} -- given in
eq.\refb{heltr} -- 
which can be defined everywhere
in the moduli space. This receives contribution from only three string
junctions supported on three D3-branes, and the corresponding walls of
marginal stability have simple structure, dividing the moduli space into
two regions -- the one where the state exists and the one where the state
does not exist. Using the wall crossing formula for this index derived 
earlier
in the context of $\NN=4$ supersymmetric string 
theories\cite{0802.1556,0803.3857} we 
derive in eq.\refb{ejump3} a simple 
formula for $B_6$ in the region where the state exists.

The other two indices $B_2(z)$ and $B_1(y,z)$,
which we introduce in 
eqs.\refb{edefb2z} and \refb{edefb1yz},
do not exist everywhere in the moduli space but
can be defined when only two of the six adjoint 
Higgs fields of the gauge theory
take vacuum expectation values. Equivalently this corresponds to a 
configuration of D3-branes where all the D3-branes lie in a plane.
On such subspaces of the moduli space the gauge theory has an unbroken
$SO(4)\equiv SU(2)_L\times SU(2)_R$ R-symmetry, and the indices
$B_2(z)$ and $B_1(y,z)$
are twisted indices which keep track of the R-charges and
the angular momentum of the system. 
These receive contribution from all planar string network with ends
lying on the D3-branes and have more complicated structure of the
walls of marginal stability compared to the three pronged string which
contributes to the sixth helicity trace.
For half BPS states these indices are easy to compute and are given in
eq.\refb{eindexhalfbps}.
We 
give wall crossing formul\ae\
for these indices in \S\ref{s3}
following the physical derivation of the Kontsevich-Soibelman (KS)
formula\cite{0811.2435,0910.4315,1006.2706} 
given in \cite{1011.1258}. 
As in the case of KS formula, the wall crossing formula is the statement
that the quantities given in eqs.\refb{eksfor}, \refb{eksformot} 
remain unchanged as we cross a wall of marginal stability.
Using these wall crossing formul\ae\ 
and the fact
that for collinear D3-brane configurations (\i.e.\ with only one adjoint Higgs field
getting vacuum expectation value) the only BPS states are the half BPS states
whose spectrum and the indices are known explicitly, we can compute the 
indices everywhere in the moduli space. We do not have a
closed form expression for these indices
since the result depends not only on the charges but
also on the region of the
moduli space we are in. However in \S\ref{sexample}
we illustrate the procedure for computing these
indices with the help of three examples.

\sectiono{Review of string network} \label{s1.5}

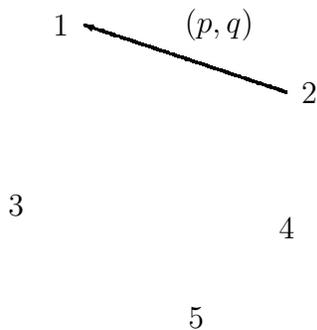
\begin{figure}
\begin{center}
\def\JPicScale{0.6}
\ifx\JPicScale\undefined\def\JPicScale{1}\fi
\unitlength \JPicScale mm
\begin{picture}(90,80)(0,0)
\linethickness{0.3mm}

\put(20,80){\makebox(0,0)[cc]{1}}

\put(10,40){\makebox(0,0)[cc]{3}}

\put(70,35){\makebox(0,0)[cc]{4}}

\put(75,65){\makebox(0,0)[cc]{2}}

\put(50,15){\makebox(0,0)[cc]{5}}

\put(55,80){\makebox(0,0)[cc]{$(p,q)$}}

\linethickness{0.3mm}
\multiput(25,80)(0.36,-0.12){125}{\line(1,0){0.36}}
\put(25,80){\vector(-3,1){0.12}}
\end{picture}

\end{center}

\vskip -32pt

\caption{A string network representation of half BPS states in
$\NN=4$ supersymmetric $SU(5)$ theory. 1,2,3,4,5
denote the positions of the five D3-branes on which the open strings can
end. Although we have displayed all the D3-branes in a plane,
each of these D3-branes can actually move along six directions transverse
to the D3-brane, representing the vacuum expectation values of the
six higgs fields in the adjoint representation of the gauge group. 
 \label{f1}}
\end{figure}

\begin{figure}


\begin{center}
\def\JPicScale{1.0}
\ifx\JPicScale\undefined\def\JPicScale{1}\fi
\unitlength \JPicScale mm
\begin{picture}(90,80)(0,0)
\linethickness{0.3mm}

\linethickness{0.3mm}
\multiput(25,80)(0.12,-0.18){167}{\line(0,-1){0.18}}
\linethickness{0.3mm}
\multiput(40,40)(0.12,0.24){42}{\line(0,1){0.24}}
\linethickness{0.3mm}
\multiput(40,40)(0.12,-0.12){83}{\line(1,0){0.12}}
\linethickness{0.3mm}
\multiput(45,20)(0.12,0.24){42}{\line(0,1){0.24}}
\linethickness{0.3mm}
\multiput(50,30)(0.24,0.12){83}{\line(1,0){0.24}}
\put(20,80){\makebox(0,0)[cc]{1}}

\put(15,40){\makebox(0,0)[cc]{3}}

\put(75,40){\makebox(0,0)[cc]{4}}

\linethickness{0.3mm}
\multiput(45,50)(0.2,0.12){125}{\line(1,0){0.2}}
\linethickness{0.3mm}
\put(20,40){\line(1,0){20}}
\put(75,65){\makebox(0,0)[cc]{2}}

\put(40,20){\makebox(0,0)[cc]{5}}

\put(65,55){\makebox(0,0)[cc]{$(p_2,q_2)$}}

\put(20,35){\makebox(0,0)[cc]{$(p_3,q_3)$}}

\put(55,20){\makebox(0,0)[cc]{$(p_5,q_5)$}}

\put(40,75){\makebox(0,0)[cc]{$(p_1,q_1)$}}

\put(75,35){\makebox(0,0)[cc]{$(p_4,q_4)$}}

\end{picture}

\end{center}

\vskip -32pt

\caption{A string network representation of quarter BPS states in
$\NN=4$ supersymmetric $SU(5)$ theory.  
Unless otherwise labelled the label $(p,q)$ on an external string will
denote a $(p,q)$ string entering the D3-brane. The electric and the magnetic
charges carried by such a configuration are given by
$Q=(q_1,q_2,q_3,q_4,q_5)$ and $P=(p_1,p_2,p_3,p_4,p_5)$
respectively. Since $\sum_i q_i = 0 =\sum_i p_i$ we see that the
configuration is neutral under the overall $U(1)$ factor of the $U(5)$
gauge theory living on the five D3-branes. \label{f2}
}
\end{figure}
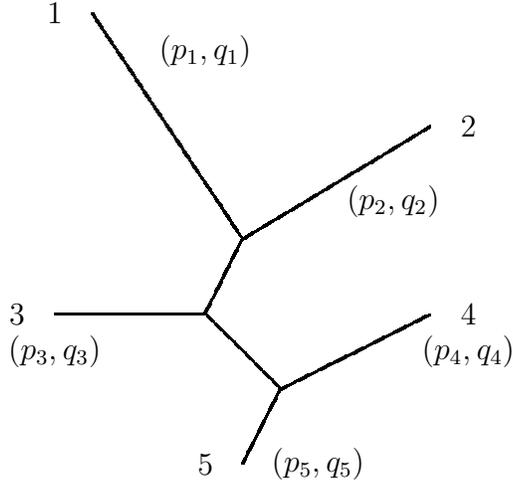

In this section we shall review some aspects of the string
network on D3-branes which represent BPS states in $\NN=4$
supersymmetric Yang-Mills theories.
We begin by reviewing the basic rules for 
constructing supersymmetric string 
network\cite{9607201,9704170,9710116,9711094,9711130,9712211,9804160,
9812021}:
\begin{enumerate}
\item The links of the network are made of $(p,q)$ 
strings, where we use the
convention that a (1,0) string represents a D-string and a
(0,1) string represents a fundamental  string. Each such string
must end either on an external D3-brane or on an internal 3-string
vertex.
\item
The $(p,q)$ values
associated with
a given external or internal string need
not be relatively prime. If for example $(p,q) = s(m,n)$ with
$m,n$ relatively prime, then $(p,q)$ 
represents $s$ copies of the $(m,n)$ string.
\item At any junction the sum of the $(p_i,q_i)$ charges carried
by the outgoing strings must vanish.
\item The network must be planar.
\item If $\tau$ denotes the complex coupling constant of the theory,
with its real part given by $\theta/2\pi$ and its imaginary part given by
$4\pi/g_{YM}^2$, then the $(p,q)$ string must lie along the direction
$e^{i\alpha}(p\bar\tau + q)$ in the two dimensional plane of the string 
network.\footnote{We could also consider another class of supersymmetric
network for which the $(p,q)$ string lies along $e^{i\alpha}(p\tau+q)$.
These two classes will be called respectively class A and class B
quarter BPS states in \S\ref{s1}. They differ from each other in the way their
unbroken supersymmetries transform under R-symmetry.} 
Here $\alpha$ is an arbitrary constant,
but it must take the same value for all the strings in a given network.
The charge conservation at each junction then also
guarantees that the net force on the junction due to the tensions of
different strings cancel.
\end{enumerate}
Some examples of string network on multiple D3-branes have been shown
in Figs.\ref{f1} and \ref{f2}. Fig.\,\ref{f1}, containing a single $(p,q)$
string stretched between two D3-branes, represents a half-BPS state,
whereas more general planar string networks of the kind shown
in Fig.\,\ref{f2} describe quarter BPS states.

\begin{figure}

\vskip -48pt

\begin{center}
\ifx\JPicScale\undefined\def\JPicScale{1}\fi
\unitlength \JPicScale mm
\begin{picture}(75,75)(0,0)
\linethickness{0.3mm}
\multiput(10,70)(0.12,-0.12){167}{\line(1,0){0.12}}
\linethickness{0.3mm}
\put(30,50){\line(1,0){30}}
\linethickness{0.3mm}
\put(60,30){\line(0,1){20}}
\linethickness{0.3mm}
\put(30,30){\line(1,0){30}}
\linethickness{0.3mm}
\put(30,30){\line(0,1){20}}
\linethickness{0.3mm}
\multiput(60,50)(0.12,0.12){83}{\line(1,0){0.12}}
\linethickness{0.3mm}
\multiput(60,30)(0.12,-0.12){125}{\line(1,0){0.12}}
\linethickness{0.3mm}
\multiput(10,10)(0.12,0.12){167}{\line(1,0){0.12}}
\linethickness{0.3mm}
\multiput(25,55)(1.95,0){21}{\line(1,0){0.98}}
\linethickness{0.3mm}
\multiput(65,25)(0,1.94){16}{\line(0,1){0.97}}
\linethickness{0.3mm}
\multiput(25,25)(1.95,0){21}{\line(1,0){0.98}}
\linethickness{0.3mm}
\multiput(25,25)(0,1.94){16}{\line(0,1){0.97}}
\put(10,75){\makebox(0,0)[cc]{1}}

\put(75,65){\makebox(0,0)[cc]{2}}

\put(5,5){\makebox(0,0)[cc]{3}}

\put(75,10){\makebox(0,0)[cc]{4}}

\linethickness{0.55mm}
\multiput(10,70)(1.43,-1.43){11}{\multiput(0,0)(0.12,-0.12){6}{\line(1,0){0.12}}}
\linethickness{0.75mm}
\multiput(65,25)(1.33,-1.33){8}{\multiput(0,0)(0.11,-0.11){6}{\line(1,0){0.11}}}
\linethickness{0.6mm}
\multiput(65,55)(1.43,1.43){4}{\multiput(0,0)(0.12,0.12){6}{\line(1,0){0.12}}}
\linethickness{0.6mm}
\multiput(10,10)(1.43,1.43){11}{\multiput(0,0)(0.12,0.12){6}{\line(1,0){0.12}}}
\end{picture}

\end{center}

\vskip -32pt

\caption{Changing the size of an internal face in a string network. The solid lines
represent the initial configuration and the dashed lines the deformed configuration.
\label{f1.5}}
\end{figure}
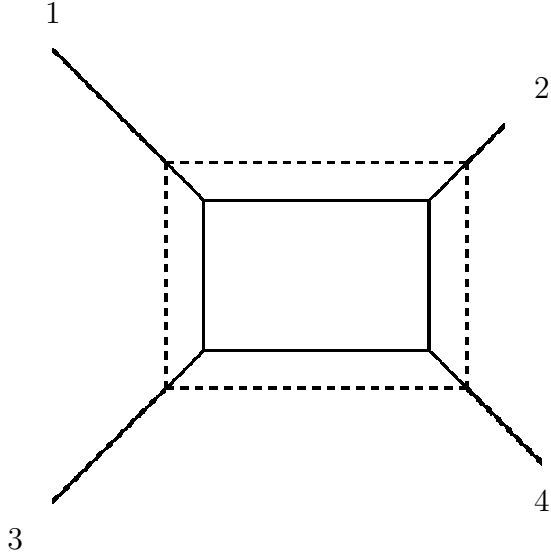

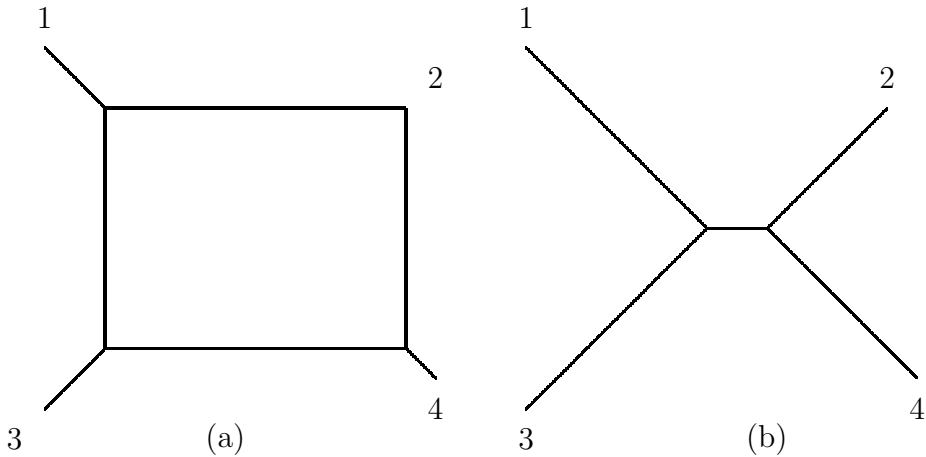
\begin{figure}

\vskip 0pt

\begin{center}
\def\JPicScale{0.8}
\ifx\JPicScale\undefined\def\JPicScale{1}\fi
\unitlength \JPicScale mm
\begin{picture}(155,75)(0,0)
\linethickness{0.3mm}
\multiput(10,70)(0.12,-0.12){83}{\line(1,0){0.12}}
\linethickness{0.3mm}
\multiput(70,20)(0.12,-0.12){42}{\line(1,0){0.12}}
\linethickness{0.3mm}
\multiput(10,10)(0.12,0.12){83}{\line(1,0){0.12}}
\put(10,75){\makebox(0,0)[cc]{1}}

\put(75,65){\makebox(0,0)[cc]{2}}

\put(5,5){\makebox(0,0)[cc]{3}}

\put(75,10){\makebox(0,0)[cc]{4}}

\linethickness{0.3mm}
\put(20,60){\line(1,0){50}}
\linethickness{0.3mm}
\put(70,20){\line(0,1){40}}
\linethickness{0.3mm}
\put(20,20){\line(1,0){50}}
\linethickness{0.3mm}
\put(20,20){\line(0,1){40}}
\linethickness{0.3mm}
\multiput(90,70)(0.12,-0.12){250}{\line(1,0){0.12}}
\linethickness{0.3mm}
\multiput(90,10)(0.12,0.12){250}{\line(1,0){0.12}}
\linethickness{0.3mm}
\multiput(130,40)(0.12,0.12){167}{\line(1,0){0.12}}
\linethickness{0.3mm}
\multiput(130,40)(0.12,-0.12){208}{\line(1,0){0.12}}
\linethickness{0.3mm}
\put(120,40){\line(1,0){10}}
\put(90,75){\makebox(0,0)[cc]{1}}

\put(150,65){\makebox(0,0)[cc]{2}}

\put(90,5){\makebox(0,0)[cc]{3}}

\put(155,10){\makebox(0,0)[cc]{4}}

\put(40,5){\makebox(0,0)[cc]{(a)}}

\put(130,5){\makebox(0,0)[cc]{(b)}}

\end{picture}

\end{center}

\vskip -24pt

\caption{Boundaries of the deformations of the internal face 
shown in Fig.\,\ref{f1.5}.
\label{f2.5}}
\end{figure}

Although we have drawn the network in Fig.\,\ref{f2}
with the topology of  a tree
there can also be networks with internal faces\cite{9804160}.
For each such face the network possesses a bosonic zero mode
that corresponds to changing the size of the face by parallel transport
of the strings which form the
perimeter of the face (see Fig.\,\ref{f1.5}). 
The general rule for such deformations is that a $(p_i,q_i)$ string along
the perimeter is moved by $\eps \, e^{i\alpha} (p_i - q_i\bar\tau)/(p_i^2 + q_i^2)$
for some small real number $\eps$. For deformations associated
with a given face $\eps$ is constant for
all the edges, but for different faces $\eps$ can be chosen differently.
Such deformations can hit
boundaries
when the face hits an external D3-brane (see Fig.\,\ref{f2.5}(a))
or shrinks to zero size (see Fig.\,\ref{f2.5}(b)).

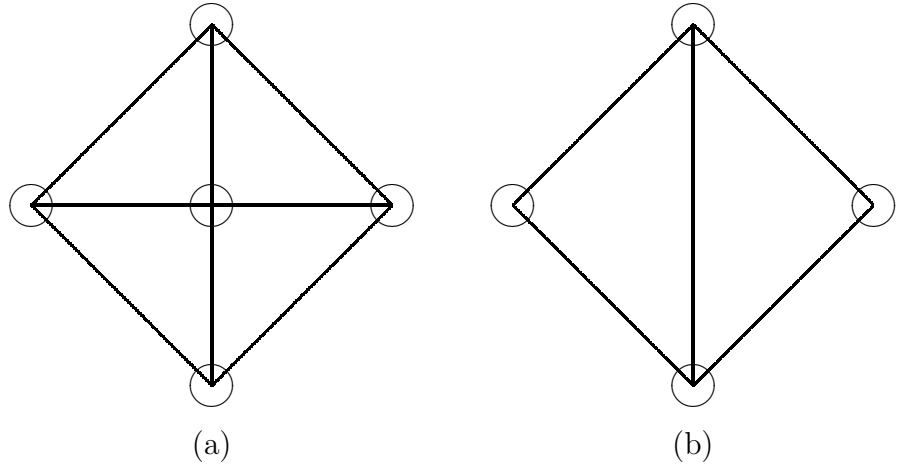
\begin{figure}

\vskip 0pt

\begin{center}
\def\JPicScale{0.8}
\ifx\JPicScale\undefined\def\JPicScale{1}\fi
\unitlength \JPicScale mm
\begin{picture}(153.54,73.54)(0,0)
\linethickness{0.3mm}
\multiput(10,40)(0.12,0.12){250}{\line(1,0){0.12}}
\linethickness{0.3mm}
\multiput(40,70)(0.12,-0.12){250}{\line(1,0){0.12}}
\linethickness{0.3mm}
\multiput(40,10)(0.12,0.12){250}{\line(1,0){0.12}}
\linethickness{0.3mm}
\multiput(10,40)(0.12,-0.12){250}{\line(1,0){0.12}}
\linethickness{0.3mm}
\put(10,40){\line(1,0){30}}
\linethickness{0.3mm}
\put(40,40){\line(0,1){30}}
\linethickness{0.3mm}
\put(40,40){\line(1,0){30}}
\linethickness{0.3mm}
\put(40,10){\line(0,1){30}}
\linethickness{0.3mm}
\multiput(90,40)(0.12,0.12){250}{\line(1,0){0.12}}
\linethickness{0.3mm}
\multiput(120,70)(0.12,-0.12){250}{\line(1,0){0.12}}
\linethickness{0.3mm}
\multiput(120,10)(0.12,0.12){250}{\line(1,0){0.12}}
\linethickness{0.3mm}
\multiput(90,40)(0.12,-0.12){250}{\line(1,0){0.12}}
\put(40,0){\makebox(0,0)[cc]{(a)}}

\put(120,0){\makebox(0,0)[cc]{(b)}}

\linethickness{0.3mm}
\put(40,70){\circle{7.07}}

\linethickness{0.3mm}
\put(70,40){\circle{7.07}}

\linethickness{0.3mm}
\put(10,40){\circle{7.07}}

\linethickness{0.3mm}
\put(40,10){\circle{7.07}}

\linethickness{0.3mm}
\put(40,40){\circle{7.07}}

\linethickness{0.3mm}
\put(120,70){\circle{7.07}}

\linethickness{0.3mm}
\put(90,40){\circle{7.07}}

\linethickness{0.3mm}
\put(150,40){\circle{7.07}}

\linethickness{0.3mm}
\put(120,10){\circle{7.07}}

\linethickness{0.3mm}
\put(120,10){\line(0,1){60}}
\end{picture}

\end{center}

\vskip 0pt

\caption{Dual grid diagrams corresponding to the string network of
(a) Fig.\,\ref{f1.5} and (b) Fig.\,\ref{f2.5}(b). The circles represent the vertices
of the grid dual to the faces of the original diagram.
\label{f3.5}
}
\end{figure}

A convenient way of representing a string network is the dual grid
diagram\cite{9710116,9804160} in which the faces are represented as 
vertices, vertices
are represented as faces, and the links are represented
as links. The precise rule for drawing the grid
diagram is as follows. We take any face of the original diagram and
declare it as the origin of the dual diagram. 
Then the other vertices in the dual grid diagram 
are chosen such that if two adjacent faces in the
original network are separated by a $(p,q)$ string then the corresponding
vertices in the dual diagram are separated by a vector $(q,-p)$.
Since at each vertex of the original
diagram we have charge conservation, this guarantees that in the dual diagram
the links forming the boundary of the face close. Fig.\,\ref{f3.5}(a) shows the
grid diagram dual to the string network of Fig.\,\ref{f1.5}. Note that the internal
points in a grid diagram represent internal faces in the original diagram. 
The grid diagram
remains invariant under the deformation described in Fig.\,\ref{f1.5}, but when the
internal face shrinks to zero size, as in Fig.\,\ref{f2.5}(b), in the dual grid
diagram the lines ending at the corresponding
vertex gets removed. For example Fig.\,\ref{f3.5}(b) shows the
grid diagram corresponding to the string network shown in Fig.\,\ref{f2.5}(b).
Conversely, existence of an integral lattice point in the interior of the
grid implies that the network admits a deformation where an internal face grows
and the corresponding dual grid diagram would correspond to connecting the
internal lattice point to its neighbors by links.

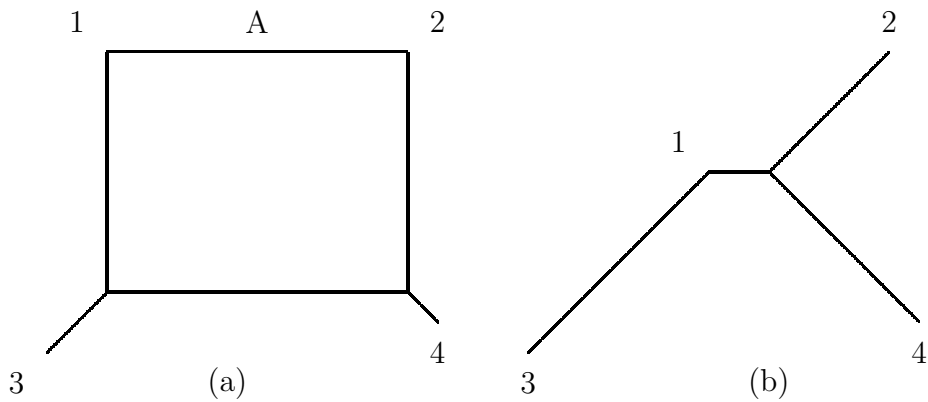
\begin{figure}

\vskip 0pt

\begin{center}
\def\JPicScale{0.8}
\ifx\JPicScale\undefined\def\JPicScale{1}\fi
\unitlength \JPicScale mm
\begin{picture}(155,65)(0,0)
\linethickness{0.3mm}
\multiput(70,20)(0.12,-0.12){42}{\line(1,0){0.12}}
\linethickness{0.3mm}
\multiput(10,10)(0.12,0.12){83}{\line(1,0){0.12}}
\put(15,65){\makebox(0,0)[cc]{1}}

\put(75,65){\makebox(0,0)[cc]{2}}

\put(5,5){\makebox(0,0)[cc]{3}}

\put(75,10){\makebox(0,0)[cc]{4}}

\linethickness{0.3mm}
\put(20,60){\line(1,0){50}}
\linethickness{0.3mm}
\put(70,20){\line(0,1){40}}
\linethickness{0.3mm}
\put(20,20){\line(1,0){50}}
\linethickness{0.3mm}
\put(20,20){\line(0,1){40}}
\linethickness{0.3mm}
\multiput(90,10)(0.12,0.12){250}{\line(1,0){0.12}}
\linethickness{0.3mm}
\multiput(130,40)(0.12,0.12){167}{\line(1,0){0.12}}
\linethickness{0.3mm}
\multiput(130,40)(0.12,-0.12){208}{\line(1,0){0.12}}
\linethickness{0.3mm}
\put(120,40){\line(1,0){10}}
\put(115,45){\makebox(0,0)[cc]{1}}

\put(150,65){\makebox(0,0)[cc]{2}}

\put(90,5){\makebox(0,0)[cc]{3}}

\put(155,10){\makebox(0,0)[cc]{4}}

\put(40,5){\makebox(0,0)[cc]{(a)}}

\put(130,5){\makebox(0,0)[cc]{(b)}}

\put(45,65){\makebox(0,0)[cc]{A}}

\end{picture}

\end{center}

\vskip 0pt

\caption{Marginal stability walls
for the string networks displayed in Fig.\,\ref{f2.5}.
\label{f4.5}
}
\end{figure}

For a given network characterized by the charges carried by the
external strings one can move around in the moduli space of the theory by moving
the positions of the D3-branes. As long as the D3-branes all lie
in a plane the network remains planar and one can preserve the BPS 
nature of the network. During such movements of the moduli one can
hit walls of marginal stability along which the original network becomes marginally
unstable against decay into two or more smaller networks carrying the same total
mass and charge. Typically this happens as one or more of the external strings
shrink to zero size. We have shown in Fig.\,\ref{f4.5} two such examples, both involving the
networks shown in Fig.\,\ref{f2.5}. 
In the first example the original network displayed in
Fig.\,\ref{f2.5}(a) becomes unstable
against decay into a single string (labelled by A) 
stretched from 1 to 2 and the rest of the network. In the second example the
network becomes unstable against decay into a single string stretched
between 1 and 3 and the rest of the 
network ending on the D3-branes 1, 2 and 4.

For identifying all the walls of marginal stability correctly it is always
best to work with the fully deformed diagram where all possible
internal faces have finite size. In the dual grid
diagram this will require that
each internal point is connected to its neighbors. The networks
where some of the internal faces shrink to zero size can be regarded as special
points in the deformation space of this more general network
parametrized by the bosonic zero modes. Conversely, given any network
with internal faces, we can associate with it a tree graph where 
all internal faces have
been shrunk to zero size. Thus the necessary condition for a network
to exist in some chamber of the moduli space is that the its associated
tree must exist in the same chamber.

Finally we note that when all the D3-branes are along a line, the only string
networks which exist are single $(p,q)$ strings stretched between a pair
of D3-branes. Any more complicated configuration can be ruled out as
follows. First of all we note that the string network must lie along the
same line along which the D3-branes lie, since 
otherwise the vertex in the network farthest from this line
will be pulled towards this line by all the strings ending at the
vertex and such a system cannot be in equilibrium. Thus all the strings in the
network must be collinear to the line along which the D3-branes lie. 
Since the relative orientation of
different strings are fixed by the charges they carry, the strings can be
collinear iff they carry parallel $(p,q)$ charges. In particular the strings
entering or leaving different D3-branes must also carry parallel $(p,q)$
charges. This means that the total electric and magnetic charges
carried by the network are parallel. Such a configuration can be rotated
to purely electrically charged configuration using S-duality and the spectrum
is that of half-BPS W-bosons of the theory, represented by single $(0,1)$
strings stretched between pairs of D3-branes. 
After reversing the S-duality
transformations they correspond to single $(p,q)$ strings stretched between
a pair of D3-branes with $(p,q)$ relatively prime.

\sectiono{Three indices}  \label{s1}

Suppose we have a BPS state that breaks $2n$ supersymmetries. Then 
there will be $2n$ fermion zero modes (goldstinos) on the world-line 
of the state. To see the effect of these zero modes consider 
a pair of fermion zero modes $\bf \psi_0$, $\bf \psi_0^\dagger$
satisfying
\be \label{ess1} 
\{ \psi_0, \psi_0^\dagger\} = 1\, .
\ee
Let us denote by $J_3$ the third component of the angular momentum and
suppose that we have chosen the basis of zero modes such that $\psi_0$
has $J_3=-1/2$ and $\psi_0^\dagger$ has $J_3=1/2$.
If $|0\rangle$ is the state annihilated by $\psi_0$ then
$|0\rangle$ and $\psi_0^\dagger |0\rangle$ will carry $J_3$ eigenvalues
$-1/4$ and $1/4$ respectively. Thus we have
\be \label{ess2}
{\rm {\rm Tr\,}} e^{2i\pi J_3} = 0, \qquad {\rm {\rm Tr\,}}e^{2i\pi J_3} (2J_3) = i
\ee
Thus the usual Witten index
${\rm Tr\,}(-1)^F={\rm Tr\,} e^{2i\pi J_3}$
will receive vanishing contribution from this sector reflecting the
fact that the quantization of the fermion zero modes
produces equal number
of bosonic and fermionic states.
To remedy this situation, we define a new index called the helicity
trace index\cite{9611205,9708062}:
\be \label{heltr}
B_{n} = {(-i)^n\over n!}\, {\rm Tr\,}\lbrace e^{2i\pi J_3} (2J_3)^{n}\rbrace \, .
\ee
The trace is taken over states carrying 
a fixed set of charges.
To see how this solves the problem let us denote by $J_3^{(1)},
\cdots J_3^{(n)}$ the contribution to $J_3$ from the $n$ pairs of
fermion zero modes and by $J_3^{\rm rest}$ the contribution to
$J_3$ from the rest of the degrees of freedom. Then we have
\be \label{ess3}
B_{n} = {(-i)^n\over n!} {\rm Tr\,}_{\rm rest} {\rm Tr\,}_{\rm zero} \left\{
e^{2i\pi\left\{J^{(1)}_3+\cdots J^{(n)}_3 + J^{\rm rest}_3\right\}}
\left(2J^{(1)}_3+\cdots 2J^{(n)}_3 + 2J^{\rm rest}_3\right)^{n}\right\}\, .
\ee
For every pair of fermion zero modes, ${\rm Tr\,} \lbrace
e^{2i\pi J_3^{(i)}}\rbrace$
vanishes but
${\rm Tr\,}\lbrace e^{2i\pi J_3^{(i)}} (2J^{(i)}_3)\rbrace$ gives a
 non-vanishing result $i$. 
Thus the only non-vanishing contribution to \refb{ess3}
comes from the
term $n! \, 2J^{(1)}_3\times 2J^{(2)}_3 \times \cdots 2J^{(n)}_3$
in the binomial expansion of 
$\left(2J^{(1)}_3+\cdots 2J^{(n)}_3 + 2J^{\rm rest}_3\right)^{n}$. 
For this term the trace over the fermion
zero modes gives a contribution of $i^n n!$.
Cancelling this against the explicit factor
of $(-i)^n / n!$ included in the definition of $B_n$ we are left with
\be \label{ess4}
B_n = {\rm Tr\,}_{\rm rest} \left\{e^{2i\pi J_3^{\rm rest}}\right\}\, .
\ee
This is in general non-vanishing.
On the other
hand, any state that breaks more than 
$2n$ supersymmetries will have more
then $n$ pairs of fermion zero modes and 
will give vanishing contribution
to this trace. In particular, non-BPS states will not contribute.
This shows that the index cannot change
under a continuous change in the moduli and hence is
protected from quantum corrections. It can however change discontinuously
across
the walls of marginal 
stability which will be discussed
later.  

We can generalize this construction as follows. Suppose that the
theory has a global symmetry $g$ under which $2m$ of the broken
supersymmetries and certain number of unbroken 
supersymmetries of the BPS state are invariant. Then it follows from
the argument given above that the index
\be \label{ebgn}
B^g_m =  {(-i)^m\over m!}\, {\rm Tr\,}\lbrace e^{2i\pi J_3} \, g \, (2J_3)^{m}\rbrace \, ,
\ee
is protected, and is in general non-zero. 
Such an index contains information about the $g$ quantum
numbers of the BPS states.

Let us now apply these general considerations to $\NN=4$
supersymmetric Yang-Mills theories in four dimensions. 
This theory has 16 supersymmetries.
Thus half BPS states break 8 supersymmetries and the relevant index is
$B_4$. 
As we have already mentioned, the result for $B_4$ is known completely.
In particular for the configuration displayed in Fig.\,~\ref{f1} we have
\be \label{eb4}
B_4(p,q) = \cases{\hbox{1 for gcd($q,p$)=1}\cr
\hbox{0 otherwise}}\, .
\ee

\begin{figure}

\vskip -24pt

\begin{center}
\def\JPicScale{0.8}
\ifx\JPicScale\undefined\def\JPicScale{1}\fi
\unitlength \JPicScale mm
\begin{picture}(90,80)(0,0)
\linethickness{0.3mm}

\put(20,80){\makebox(0,0)[cc]{1}}

\put(15,45){\makebox(0,0)[cc]{3}}

\put(70,35){\makebox(0,0)[cc]{4}}

\put(75,65){\makebox(0,0)[cc]{2}}

\put(45,20){\makebox(0,0)[cc]{5}}

\linethickness{0.3mm}
\multiput(25,75)(0.12,-0.18){83}{\line(0,-1){0.18}}
\linethickness{0.3mm}
\multiput(35,60)(0.83,0.12){42}{\line(1,0){0.83}}
\linethickness{0.3mm}
\multiput(20,45)(0.12,0.12){125}{\line(1,0){0.12}}
\end{picture}

\end{center}

\vskip -48pt

\caption{A string network representation of quarter BPS states in
$\NN=4$ supersymmetric $SU(5)$ theory with three external strings ending on
three D3-branes. Such configurations contribute to $B_6$.
\label{f3}}
\end{figure}
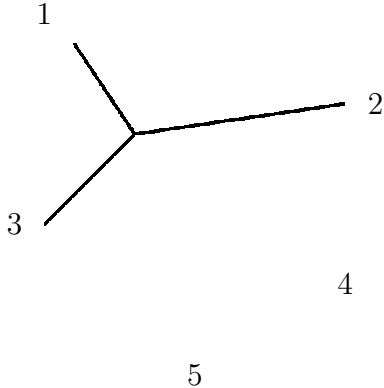

Quarter BPS states break 12 supersymmetries and
the relevant index for counting these states is $B_6$.
However the only configurations which contribute
to $B_6$ are those 
containing three external strings ending on three D3-branes, {\it e.g.}
the one shown in
Fig.\,~\ref{f3}. The reason for this is that only planar string networks describe
BPS configurations. Since we can always draw a plane through three points,
any configuration that has ends on at most three D3-branes can always be
made planar. In contrast a more general configuration, like the one
shown in Fig.\,~\ref{f2}, is necessarily non-planar  if the D3-brane
coordinates do not lie in a plane.
Such a
configuration is non-BPS and hence does not contribute to $B_6$. Since 
we have argued that $B_6$ is invariant under continuous deformation
of the moduli, it follows that even when all the D3-branes lie
in a plane, planar networks ending on four or more D3-branes must
have vanishing $B_6$. Physically this has its origin in the fact that such
planar networks have some additional fermion zero modes besides the
ones associated with the 12 broken supersymmetries\cite{9804160}, 
and the trace
over these fermion zero modes makes the index vanish.

\begin{table}
\begin{center}
\def\st{\vrule height 3ex width 0ex}
\begin{tabular}{|l|l|l|l|l|l|l|l|l|l|l|} \hline
BPS state & unbroken supersymmetries & broken supersymmetries  
\st\\[1ex] \hline\hline
Half BPS & (1,2,2) + (2,1,2) & (1,2,2) + (2,1,2)  \st\\[1ex] \hline
Class A quarter BPS & (1,2,2) &  (1,2,2) + 2 (2,1,2) \st\\[1ex] \hline
Class B quarter BPS & (2,1,2) & 2(1,2,2) + (2,1,2)  \st\\[1ex] \hline\hline
\end{tabular}
\caption{$SU(2)_L\times SU(2)_R\times SU(2)_{\rm rotation}$
transformation laws of various supersymmetries. \label{t1}}
\end{center}
\end{table}

Are there other protected indices which can capture information about
the BPS states associated with the planar string network? From the
discussion above it should be clear that any index that can be defined at
a generic point in the moduli space must vanish for planar networks with
four or more external strings since in that case we can
compute that index by going to a non-planar configuration of D3-branes
where the state is manifestly non-supersymmetric and hence gives
vanishing contribution to the index. Thus we need to look for indices
which are defined only for planar configuration of the 
D3-branes.\footnote{Twisted indices which are defined on a subspace
of the full moduli space played an important role in testing the
correspondence between black holes and microstates at the
non-perturbative level\cite{0911.1563,1002.3857}.
At the level of supersymmetric quantum mechanics describing the
dynamics of multiple monopoles in $\NN=4$ supersymmetric Yang-Mills
theories, such indices have been introduced by Stern and Yi\cite{0005275}.
}
To this
end we note that when all the D3-branes lie in a plane, the theory has an
additional unbroken $SO(4)\simeq SU(2)_L\times SU(2)_R$ symmetry
corresponding to rotation in the four directions transverse to the plane of
the D3-brane. In the language of the supersymmetric Yang-Mills theory
this SO(4) symmetry is a subgroup of the SO(6) R-symmetry group that
remains unbroken when only two of the six adjoint Higgs fields acquire
vacuum expectation values. 
The quarter BPS states of the theory can be divided into two classes, which
we shall call class A and class B states, according to the transformations
properties of the unbroken supersymmetries under the $SU(2)_L\times SU(2)_R$
transformation. We have shown in table \ref{t1} 
the transformation laws of the unbroken and
broken supersymmetries under the $SU(2)_L\times SU(2)_R\times
SU(2)_{\rm rotation}$ group for different types of BPS states, with
$SU(2)_{\rm rotation}$ denoting the usual rotation
group in the three space dimensions.
In particular we note that for class A quarter BPS states and also half
BPS states there are four $SU(2)_L$ invariant unbroken supersymmetries,
and four $SU(2)_L$ invariant broken super symmetries. 
We shall denote by $I_{3L}$ and $I_{3R}$ the
third components of the generators of $SU(2)_L$ and $SU(2)_R$
respectively. It now follows from \refb{ebgn} that the index
\be \label{edefb2z}
B_2(z) = -{1\over 2!} {\rm Tr\,}\left\{e^{2i\pi J_3}  z^{2 I_{3L}} \, (2 J_3)^2
\right\}, \qquad z\in\CCC, \quad
|z|=1\, ,
\ee
is protected and 
will receive contribution from half BPS states and the class A quarter
BPS states. We can also introduce another index by replacing $I_{3L}$
by $I_{3R}$ in \refb{edefb2z} which will receive contribution from half
BPS and class B quarter BPS states but for definiteness we shall
concentrate on the index given in \refb{edefb2z}. Since this index
is defined only for planar configuration of D3-branes, it can receive
contribution from general string network ending on arbitrary number of
D3-branes.

Table \ref{t1} also shows that for half BPS and class A quarter BPS states
2 of the 
$SU(2)_L$ invariant unbroken generators and 2 of the
$SU(2)_L$ invariant broken generators are invariant under $(I_{3R}+J_3)$.
This allows us to to define yet another protected index
\be \label{edefb1yz}
B_1(y,z) = - {1\over y - y^{-1}}\,
{\rm Tr\,}\left\{e^{2i\pi J_3} z^{2 I_{3L}} \, y^{2I_{3R} + 2 J_3}
\, (2 J_3)\right\}\, , \qquad y,z\in \CCC, \quad
|z|=1\, , \quad |y|=1\, \, .
\ee
The index \refb{edefb1yz} is analogous
to the protected spin character defined in \cite{1006.0146} (with the
replacement $y\to -y$).

We have adjusted the normalizations of $B_2(z)$ and
$B_1(y,z)$ such that after factoring
our the contribution from the $SU(2)_L$ invariant fermion zero modes in the
trace we are left with 
${\rm Tr}'_{\rm rest}\left\{e^{2i\pi J_3}  z^{2 I_{3L}} 
\right\}$ and 
${\rm Tr}'_{\rm rest}\left\{e^{2i\pi J_3}  z^{2 I_{3L}} 
\, y^{2I_{3R} + 2 J_3}
\right\}$ respectively as in \refb{ess4}. The prime on Tr denotes that
it includes traces over the $SU(2)_L$ non-invariant fermion zero
modes. Thus we have the relation
\be \label{eb2b1rel}
B_2(z) =\lim_{y\to 1} B_1(y, z)\, .
\ee
We can also find a relation between $B_2(z)$ and $B_6$ by factoring out
the contribution from $SU(2)_L$ non-invariant fermion zero modes from
$B_2(z)$. From table \ref{t1} we see that for class A quarter BPS
states these transform in two (2,1,2)
representation of $SU(2)_L \times SU(2)_R \times SU(2)_{\rm rotation}$.
Their contribution to $B_2(z)$ corresponds to a factor of $(z+z^{-1}-2)^2$.
Thus $B_2(z)$ can be expressed as 
$(z+z^{-1}-2)^2{\rm Tr}_{\rm rest}\left\{e^{2i\pi J_3}  z^{2 I_{3L}} 
\, \right\}$. Comparing this with \refb{ess4} we get
\be \label{eb2b6}
B_6 = \lim_{z\to 1} \, (z+z^{-1}-2)^{-2}\,  B_2(z)\, .
\ee

For a half BPS state represented by
a $(p,q)$ string stretched between two D3-branes with $(p,q)$ relatively
prime, the contribution to $B_2(z)$ and $B_1(y,z)$ can be evaluated by
knowing the $(I_{3L}, I_{3R}, J_3)$ quantum numbers carried by the
8 fermion zero modes associated with broken supersymmetries. The
$(I_{3L}, I_{3R}, J_3)$ assignments are as follows:
\ben \label{eassign}
&& (0, 1/2, 1/2), \, (0, 1/2, -1/2), \, (0, -1/2, 1/2), \, (0, -1/2, -1/2), \nn &&
(1/2, 0, 1/2), \, (1/2, 0, -1/2), \, (-1/2, 0, 1/2), \, (-1/2, 0, -1/2)\, .
\een
Upon quantization of these zero modes we get the following
contribution to the indices $B_2(z)$ and $B_1(y,z)$:
\ben \label{eindexhalfbps}
B_2(z) &=& (z + z^{-1}-2)\, ,
\nn
B_1(y,z) &=& (z+z^{-1}-y-y^{-1})
\, .
\een

In the next section we shall describe the wall crossing formula 
for the jump in
these  indices across walls of marginal stability.
Since we have argued at the end of \S\ref{s1.5} that for collinear
configuration of
D3-branes  the only surviving BPS configurations
are the half BPS states, 
the computation of the various indices for such configurations
is straightforward. The  values of the indices elsewhere in the moduli
space can then be determined using the wall crossing formul\ae\ for the
various indices.

\sectiono{Wall crossing formul\ae} \label{swall}

In this section we shall describe the wall crossing formul\ae\ for the
three indices introduced in \S\ref{s1}.

\subsection{Sixth helicity trace index} \label{s2}

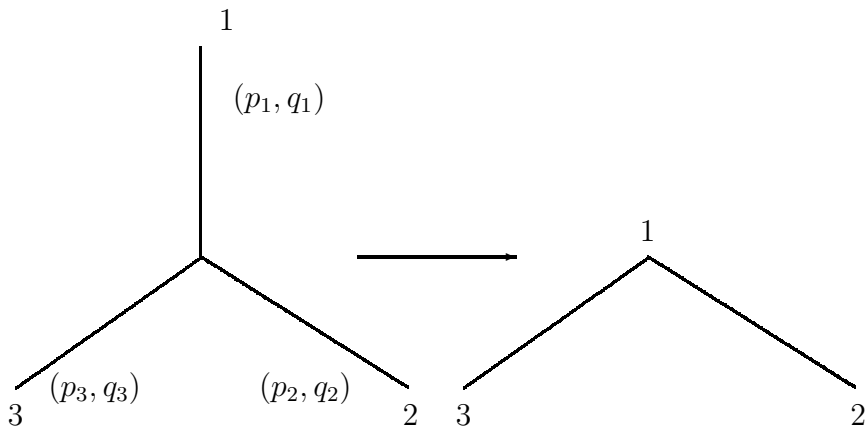
\begin{figure}
\begin{center}
\def\JPicScale{0.7}
\ifx\JPicScale\undefined\def\JPicScale{1}\fi
\unitlength \JPicScale mm
\begin{picture}(165,85)(0,0)
\linethickness{0.3mm}
\put(40,40){\line(0,1){40}}
\linethickness{0.3mm}
\multiput(40,40)(0.19,-0.12){208}{\line(1,0){0.19}}
\put(45,85){\makebox(0,0)[cc]{1}}

\put(80,10){\makebox(0,0)[cc]{2}}

\put(5,10){\makebox(0,0)[cc]{3}}

\linethickness{0.3mm}
\multiput(125,40)(0.19,-0.12){208}{\line(1,0){0.19}}
\put(55,70){\makebox(0,0)[cc]{$(p_1,q_1)$}}

\put(60,15){\makebox(0,0)[cc]{$(p_2,q_2)$}}

\put(20,15){\makebox(0,0)[cc]{$(p_3,q_3)$}}

\linethickness{0.3mm}
\multiput(5,15)(0.17,0.12){208}{\line(1,0){0.17}}
\linethickness{0.3mm}
\multiput(90,15)(0.17,0.12){208}{\line(1,0){0.17}}
\linethickness{0.3mm}
\put(70,40){\line(1,0){30}}
\put(100,40){\vector(1,0){0.12}}
\put(125,45){\makebox(0,0)[cc]{1}}

\put(165,10){\makebox(0,0)[cc]{2}}

\put(90,10){\makebox(0,0)[cc]{3}}

\end{picture}

\end{center}\caption{A marginal stability wall of the string network 
shown in Fig.\,\ref{f3}. The decay products are a $(p_2, q_2)$ string
stretched from 1 to 2, and a $(p_3, q_3)$ string stretched from
1 to 3.
\label{f4}}
\end{figure}

As argued in \S\ref{s1}, the sixth helicity trace $B_6$ receives
contribution only from configurations with three external strings. The
walls of marginal stability across which $B_6$ jumps are those on which
the state becomes unstable against decay into a pair of half BPS states.
In terms of string network such decays correspond to one of the
external strings shrinking to zero size, as
shown in Fig.\,\ref{f4}. If $(Q_1,P_1)$ and $(Q_2,P_2)$ denote the
(electric, magnetic) charges carried by the two half BPS states into
which the state decays, then the jump in the $B_6$ value as we
cross the wall from the side in which it does not exist to the one in
which it exists is given by\cite{0802.1556,0803.3857}:
\ben \label{ejump}
\Delta B_6(Q,P) &=& (-1)^{Q_1.P_2-Q_2.P_1+1} |Q_1.P_2-Q_2.P_1|
\sum_{L_1|(Q_1,P_1)} B_4(Q_1/L_1,P_1/L_1) \nn
&&\qquad \qquad \qquad \quad \qquad \qquad \qquad \times
\sum_{L_2|(Q_2,P_2)} B_4(Q_2/L_2,P_2/L_2)\, , 
\een
where $L_i|(Q_i,P_i)$ means that $L_i$ must be a common factor
of all components of $Q_i$ and $P_i$. 
For the decay displayed in Fig.\,\ref{f4} we have
\be \label{echarges}
P_1=(-p_2, p_2, 0), \quad Q_1 = (-q_2, q_2, 0), \quad
P_2=(-p_3, 0, p_3), \quad Q_2 = (-q_3, 0, q_3) \, .
\ee
Thus we have
\be \label{ecross}
Q_1.P_2 - Q_2. P_1 = (q_2 p_3 - q_3 p_2) \, .
\ee
Let $s_2=\gcd(p_2, q_2)$ and $s_3=\gcd(p_3, q_3)$. 
Now 
$(Q_1/L_1, P_1/L_1)$ represents a $(q_2/L_1, p_2/L_1)$
string stretched between the D3-branes 1 and 2. It follows
from \refb{eb4} that the index $B_4$ for such a state is non-zero  
iff $\gcd(q_2/L_1, p_2/L_1)=1$, \i.e.\ iff $L_1=s_2$. 
Similarly $B_4(Q_2/L_2, P_2/L_2)$
is non-vanishing iff $L_2=s_3$. Furthermore we have
$B_4(Q_1/s_2, P_1/s_2)=1$ and
$B_4(Q_2/s_3, P_2/s_3)=1$. Substituting these 
and \refb{ecross} into \refb{ejump} we get
\be \label{ejump2}
\Delta B_6(Q,P) = (-1)^{q_2p_3-q_3p_2+1} |q_2p_3-q_3p_2|
\, .
\ee
When the   D3 brane 1 crosses the wall of marginal stability
the configuration displayed in Fig.\,\ref{f4} ceases to exist and hence
$B_6$ vanishes. Thus \refb{ejump2} represents the value of $B_6$
on the side of the wall where the configuration exists, and we can write
\be \label{ejump3}
B_6(Q,P) = (-1)^{q_2p_3-q_3p_2+1} |q_2p_3-q_3p_2|
\, .
\ee
We note that this formula is symmetric under the exchange of the three
external strings as a consequence of the `conservation law'
\be \label{econs}
p_1+p_2+p_3=0, \qquad q_1+q_2+q_3=0\, ,
\ee
and hence we shall arrive at the same formula if we
apply the wall crossing across the other two walls of marginal stability
where either the $(p_2, q_2)$ string or the $(p_3,q_3)$ string shrinks to zero
size.

We should also add that we can consider string network with three external
strings and internal faces. 
Such configurations are planar and in principle could
contribute to $B_6$. However the marginal stability walls on which such a
string network breaks apart into a pair of half BPS states are always of the type
shown in Fig.\,\ref{f4} where the internal face has shrunk to zero size.
 Along other marginal stability walls where the original network
contains internal faces, at least one of the decay products will be
quarter BPS (see {\it e.g.} Fig.\,\ref{fn4})
and such decays do not contribute to jumps in 
$B_6$\cite{0707.1563,0809.1157,0903.2481}. 
Thus we can ignore them
for computation of $B_6$, although, as we shall see later, they will contribute to
jumps in $B_2(z)$ and $B_1(y,z)$.

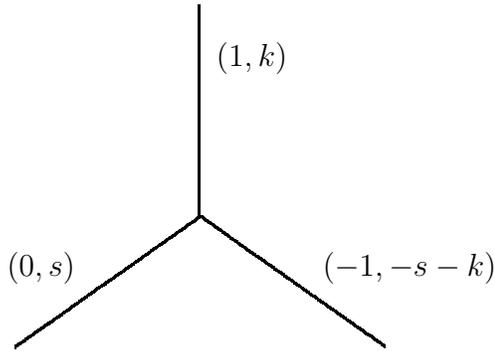
\begin{figure}

\vskip -48pt

\begin{center}
\def\JPicScale{0.7}
\ifx\JPicScale\undefined\def\JPicScale{1}\fi
\unitlength \JPicScale mm
\begin{picture}(105,90)(0,0)
\linethickness{0.3mm}
\put(70,50){\line(0,1){40}}
\linethickness{0.3mm}
\multiput(35,25)(0.17,0.12){208}{\line(1,0){0.17}}
\linethickness{0.3mm}
\multiput(70,50)(0.17,-0.12){208}{\line(1,0){0.17}}
\put(80,80){\makebox(0,0)[cc]{$(1,k)$}}

\put(40,40){\makebox(0,0)[cc]{$(0,s)$}}

\put(110,40){\makebox(0,0)[cc]{$(-1,-s-k)$}}

\end{picture}

\vskip -32pt

\caption{A special class of string network configurations.
\label{f5}
} 
\end{center}

\end{figure}

For special configurations of the type shown in Fig.\,\ref{f5} the 
expression for $B_6$ can be derived from the results of
\cite{0005275,0609055} based on the study of supersymmetric quantum mechanics
of monopole system. The result is $(-1)^{s+1} |s|$ in agreement
with \refb{ejump3}. 
For the same configuration the formula was also derived
in \cite{0802.0761} using primitive wall crossing formula. However 
for deriving the result for
most general set of charges we need to use the general wall crossing
formula for decays into non-primitive charge vectors as given in
\refb{ejump}.

\subsection{The twisted and motivic helicity trace indices}  \label{s3}

As discussed before, the index $B_6$ vanishes for planar
string network with four or more external legs but the indices
$B_2(z)$ and $B_1(y,z)$ defined in \refb{edefb2z} and \refb{edefb1yz} do not
vanish in general. However for 
collinear configuration of D3-branes these
indices do vanish except for half BPS states. 
The latter indices have been computed in \refb{eindexhalfbps}.
Thus
if we can write down the general
wall crossing formula for these indices, 
then we can compute them
at any point in the moduli space by starting with the known values of the
indices for collinear configurations and then successively applying the
wall crossing formula across each wall of marginal stability.

Since the supersymmetry subalgebra that commutes with $SU(2)_L$ is
the $\NN=2$ supersymmetry algebra, one
expects that the wall crossing formul\ae\ for $B_2(z)$ and $B_1(y,z)$ 
will 
be similar to the KS
wall crossing formula\cite{0811.2435,0910.4315,1006.2706}. 
Indeed by now there
are many physical `derivations'
 of the KS wall crossing 
 formula\cite{0807.4723,0807.4556,1006.0146,
 1008.0030,1011.1258,1102.1729,1103.1887,1107.0723,1112.2515} 
 and we can
use any of them to derive the wall crossing formula for the indices $B_2(z)$ and
$B_1(y,z)$. We have derived these formul\ae\ using the arguments given in
\cite{1011.1258,1103.1887} and the result of \cite{1112.2515} 
proving the equivalence of the wall
crossing formul\ae\ of \cite{1011.1258} and the KS wall crossing formula. 
Since the logic is identical to those in \cite{1011.1258,1112.2515}, 
we shall not give the
details of the argument but only quote the final results.

We begin by introducing some notations.
Let
$\alpha=(Q,P)$ denote the charge vector and
given two such vectors we define
\be \label{ealal}
\langle \alpha, \alpha'\rangle = Q\cdot P' - P\cdot Q'\, .
\ee
We shall denote by $Z_\gamma$ the central charge of a charge vector
$\gamma$ under the $SU(2)_L$ invariant
$\NN=2$
subalgebra. 
For a set of D3-branes at positions $z_1$, $z_2$, $\cdots$ in the complex
plane, the central charge of a planar network with $(p_i,q_i)$ string
entering the $i$-th D3-brane is given by
\be \label{eqzgamma}
Z_\gamma= {1\over \sqrt \tau_2} \sum_i \bar z_i (p_i\bar\tau + q_i)\, ,
\ee
up to a constant of proportionality.\footnote{Using the 
charge conservation at each vertex we can express
\refb{eqzgamma} as $\tau_2^{-1/2} \sum_{links\, m} \overline{\Delta z_m}
(p_m\bar\tau + q_m)$ where $\Delta z_m$ is the complex number
describing the length and orientation of the $m$-th link and $(p_m,q_m)$
is the charge of the string along the $m$-th link. Using the fact that
$\Delta z_m \propto e^{i\alpha} (p_m\bar \tau + q_m)$ we can
express this as $\tau_2^{-1/2} e^{-i\alpha}
\sum_{links\, m} |\Delta z_m| |p_m\tau + q_m|$. Since 
$|p_m\tau + q_m|/\sqrt{\tau_2}$ is the tension of the $(p_m,q_m)$ string,
we see that $|Z_\gamma|$ is
proportional to the total mass of the network as expected.}
Near any wall of marginal stability  we can find a pair of
vectors
$\gamma_1$ and $\gamma_2$ 
satisfying the following properties:
\begin{enumerate}
\item
Along the wall of marginal stability of interest
the central charges
$Z_{\gamma_1}$ and $Z_{\gamma_2}$ get aligned.
\item
Any charge vector lying in the plane of
$\gamma_1$ and $\gamma_2$ can be expressed as $m\gamma_1+
n\gamma_2$ with integer $m,n$.  
\item Near the wall of marginal stability BPS states
of charge $m\gamma_1+n\gamma_2$ exist only for $m,n\ge 0$
or $m,n\le 0$\cite{1008.0030}.
\end{enumerate}

First we shall give the wall crossing formula for $B_2(z)$
across such a wall.
We denote by 
$B_2(\alpha;z)$ the $B_2(z)$ index for charge vector
$\alpha$,
and introduce the rational index
\be \label{eratb2}
\bar B_2(\alpha; z) = \sum_{m|\alpha} m^{-2} B_2(\alpha/m; z^m)\, .
\ee
We also introduce an infinite dimensional algebra with
generators $e_\alpha$ satisfying the commutations relations:
\be \label{ecomm}
[e_\alpha, e_{\alpha'}]=(-1)^{\langle \alpha, \alpha'\rangle}
\,  \langle \alpha, \alpha'\rangle \, e_{\alpha+\alpha'}\, .
\ee
The KS wall crossing
formula is the statement that
\be \label{eksfor}
P\left(\prod_{M\ge 0,N\ge 0} \exp\left[
\bar B_2(M\gamma_1+N\gamma_2;z) e_{M\gamma_1+N\gamma_2}\right]\right)
\ee
remains unchanged across a wall of marginal stability. $P$ denotes a phase
ordered product of the exponentials such that the 
phase of $Z_{M\gamma_1+N\gamma_2}$ decreases monotonically as we
move from the left most element to the right-most element of the
product. As we cross a wall of marginal stability, the phases of $Z_{\gamma_1}$
and $Z_{\gamma_2}$ switch order and as a result the order in the
product in \refb{eksfor} is reversed. The wall crossing formula tells
us that the indices $B_2(M\gamma_1+N\gamma_2;z)$ will have to change
in such a way that the product remains invariant. Using \refb{eksfor} we can
determine the indices $B_2(\alpha;z)$ on one side of the wall of marginal
stability if we know their values on the other side.

The wall crossing formula for $B_1(\alpha;y,z)$ is a generalization of the
motivic wall crossing formula of KS.
For this we define\footnote{In the analysis of \cite{1011.1258}
the $(y-y^{-1})/(y^m - y^{-m})$ factor in \refb{emot1} arose from the fact
that for motion in a magnetic field the orbital angular momentum grows with
the magnetic field. In contrast the $SU(2)_L$ quantum number to which
$z$ couples is not affected by the magnetic field and hence there is no
such factor involving $z$.}
\be \label{emot1}
\bar B_1(\alpha; y,z) = \sum_{m|\alpha} m^{-1} {y - y^{-1}\over y^m - y^{-m}}
B_1(\alpha/m; y^m, z^m) \, ,
\ee
and introduce the infinite dimensional algebra generated by $\tilde e_{\alpha}$
satisfying the commutation relations:
\be \label{emot2}
[\tilde e_\alpha, \tilde e_{\alpha'}] = {(-y)^{\langle\alpha,\alpha'\rangle} -
(-y)^{-\langle\alpha,\alpha'\rangle} \over y-y^{-1}} \tilde e_{\alpha+\alpha'}\, .
\ee
The  wall crossing formula for $B_1(\alpha;y,z)$ then tells us that the
product
\be \label{eksformot}
P\left(\prod_{M\ge 0,N\ge 0} \exp\left[
\bar B_1(M\gamma_1+N\gamma_2;y,z) \tilde 
e_{M\gamma_1+N\gamma_2}\right]\right)
\ee
remains unchanged across the wall of marginal stability.
Using \refb{eksformot} we can
determine the indices $B_1(\alpha;y,z)$ on one side of the wall of marginal
stability if we know their values on the other side.
Note that as $y\to 1$ the wall crossing formula for $B_1(\alpha;y,z)$ tends to
that for $B_2(\alpha;z)$.

Eqs.\refb{eksfor} and \refb{eksformot} give implicit relations which
determine the index on one side in terms of the index on the other side.
Explicit formul\ae\ for the indices on one side in terms of their values
on the other side can be found in \cite{1011.1258}. The equivalence of these explicit
formul\ae\ and \refb{eksfor}, \refb{eksformot} has been proved
in \cite{1112.2515}.

Special cases of these general wall crossing formul\ae\ are the primitive
and the semi-primitive wall crossing 
formul\ae\cite{0702146}. Let us for definiteness
denote by $B_2^+$ and $B_1^+$ the indices on the side of the wall
in which
\be \label{edefplus}
\langle \gamma_1, \gamma_2\rangle \, {\rm Im} (Z_{\gamma_1}
\bar Z_{\gamma_2}) < 0\, ,
\ee
and by $B_2^-$ and $B_1^-$ the indices on the other 
side.\footnote{Physically
the $-$ side corresponds to the side in which there are
multi-centered loosely bound configurations with individual centers
carrying charges of the form $m\gamma_1+n\gamma_2$. On the + side
there are no such bound states. Hence the jump in the index can be
identified as the contribution to the index from these loosely
bound states.}
Then the primitive wall crossing formula tells us that
\ben \label{eprim}
&& B_2^-(\gamma_1+\gamma_2; z) - B_2^+(\gamma_1+\gamma_2;z)
= (-1)^{\langle \gamma_1, \gamma_2\rangle+1}\,  
|\langle \gamma_1, \gamma_2\rangle| \, 
B_2^+(\gamma_1;z) \, B_2^+(\gamma_2;z)\, , \nn
&&
B_1^-(\gamma_1+\gamma_2; y,z) - B_1^+(\gamma_1+\gamma_2;y,z)
= { (-y)^{-|\langle \gamma_1, \gamma_2\rangle|}
- (-y)^{|\langle \gamma_1, \gamma_2\rangle|}\over y - y^{-1}}
B_1^+(\gamma_1;y,z) \, B_1^+(\gamma_2;y,z)\, . \nn
\een

The semiprimitive wall crossing formula for the index $B_2(\gamma_1+
N\gamma_2;z)$ tells us that
\be \label{esemib2}
\bar B_2^-(\gamma_1+N\gamma_2; z) 
- \bar B_2^+ (\gamma_1+N\gamma_2; z) 
= \sum_{\ell=0}^{N-1} \bar B_2^+(\gamma_1 + \ell \gamma_2;z)
\, \Omega_{\rm halo}(\gamma_1, \gamma_2, N-\ell;z)\, ,
\ee
where\cite{0702146}
\be \label{eesmib3}
\sum_{N=0}^\infty \Omega_{\rm halo}(\gamma_1, \gamma_2, N;z)\, q^N
= \exp\left[-\sum_{s=1}^\infty s\, q^s (-1)^{s\langle\gamma_1,\gamma_2\rangle}
\, |\langle\gamma_1,\gamma_2\rangle|\, \bar B_2^+(s\gamma_2;z)\right]\, .
\ee
Since $\gamma_1+\ell\gamma_2$ is primitive for all $\ell$, we can replace
$\bar B_2^\pm(\gamma_1+N\gamma_2; z)$ and
$\bar B_2^+(\gamma_1 + \ell \gamma_2;z)$ in \refb{esemib2}
by
$B_2^\pm(\gamma_1+N\gamma_2; z)$ and
$B_2^+(\gamma_1 + \ell \gamma_2;z)$ respectively. On the other hand
using \refb{eratb2} and \refb{eesmib3} we get
\ben \label{ehalo2}
\sum_{N=0}^\infty \Omega_{\rm halo}(\gamma_1, \gamma_2, N;z) \, q^N
&=&
\exp\left[-\sum_{s=1}^\infty s\, q^s (-1)^{s\langle\gamma_1,\gamma_2\rangle}
\, |\langle\gamma_1,\gamma_2\rangle|\, \sum_{m|s}
m^{-2} \, B_2^+(s\gamma_2/m;z^m)\right] \nn
&=& \exp\left[-\sum_{m=1}^\infty \sum_{k=1}^\infty
m^{-1}k\, q^{mk} (-1)^{m k\langle\gamma_1,\gamma_2\rangle}
\, |\langle\gamma_1,\gamma_2\rangle|\, 
B_2^+(k\gamma_2;z^m)\right] \, . \nn
\een
If 
\be \label{eifb2}
B_2^+(k\gamma_2;z) = \sum_p B_{2,p}^+(k\gamma_2) z^p\, ,
\ee
then \refb{ehalo2} may be expressed as
\be \label{ehalo3}
\sum_{N=0}^\infty \Omega_{\rm halo}(\gamma_1, \gamma_2, N;z) \, q^N
=
\prod_{p}\prod_{k=1}^\infty \left( 1 - q^k z^p 
(-1)^{k\langle\gamma_1,\gamma_2\rangle}\right)^{k \,
|\langle\gamma_1,\gamma_2\rangle|\, B_{2,p}^+(k\gamma_2)}
\, .
\ee
This gives an expression for the change in $B_2(\gamma_1+N\gamma_2; z)$
across the wall of marginal stability.

Finally the semiprimitive wall crossing formula for the index $B_1(\gamma_1+
N\gamma_2;y,z)$ tells us that\cite{0904.1420}
\be \label{esemib1a}
\bar B_1^-(\gamma_1+N\gamma_2; y,z) 
- \bar B_1^+ (\gamma_1+N\gamma_2; y,z) 
= \sum_{\ell=0}^{N-1} \bar B_1^+(\gamma_1 + \ell \gamma_2;y,z)
\, \Omega_{\rm halo}(\gamma_1, \gamma_2, N-\ell;y,z)\, ,
\ee
where
\be \label{eesmib1bold}
\sum_{N=0}^\infty \Omega_{\rm halo}(\gamma_1, \gamma_2, N;y,z) \, q^N
= \exp\left[\sum_{s=1}^\infty \, q^s 
\, { (-y)^{-s|\langle \gamma_1, \gamma_2\rangle|}
- (-y)^{s|\langle \gamma_1, \gamma_2\rangle|}\over y - y^{-1}}\, \bar B_1^+(s\gamma_2;y,z)\right]\, .
\ee
Again using \refb{emot1} we can express \refb{eesmib1bold} as
\ben \label{eesmib1b}
&&
\sum_{N=0}^\infty \Omega_{\rm halo}(\gamma_1, \gamma_2, N;y,z) \, q^N
\nn
&=& \exp\left[\sum_{m=1}^\infty \sum_{k=1}^\infty \, m^{-1}\,  q^{mk} 
\, { (-y)^{-mk|\langle \gamma_1, \gamma_2\rangle|}
- (-y)^{mk|\langle \gamma_1, \gamma_2\rangle|}\over y^m - y^{-m}}\, 
B_1^+(k\gamma_2;y^m,z^m)\right]\, . 
\een
If
\be \label{eifb1}
B_1^+(k\gamma_2;y,z) = \sum_{n,p} B_{1,n,p}^+(k\gamma_2) y^{n} z^p\, ,
\ee
then \refb{eesmib1b} may be expressed as
\be \label{ehalo3forb1}
\sum_{N=0}^\infty \Omega_{\rm halo}(\gamma_1, \gamma_2, N;y,z) \, q^N
=
\prod_{p,n}\prod_{k=1}^\infty \prod_{r=1}
^{k|\langle \gamma_1, \gamma_2\rangle|}
\left( 1 - (-1)^{k\langle\gamma_1,\gamma_2\rangle}
q^k z^p y^{n+2r-1 - k|\langle \gamma_1, \gamma_2\rangle|}
\right)^{B_{1,n,p}^+(k\gamma_2)}
\, .
\ee

In the next section we shall see some examples of how using these
wall crossing formul\ae\ we can calculate the indices $B_2(z)$ and
$B_1(y,z)$ for planar string network.

\sectiono{Three examples} \label{sexample}

In this section we shall apply the wall crossing formul\ae\ to compute
the indices $B_2(z)$ and $B_1(y,z)$ 
of three different string network configurations.

\subsection{Example 1} 

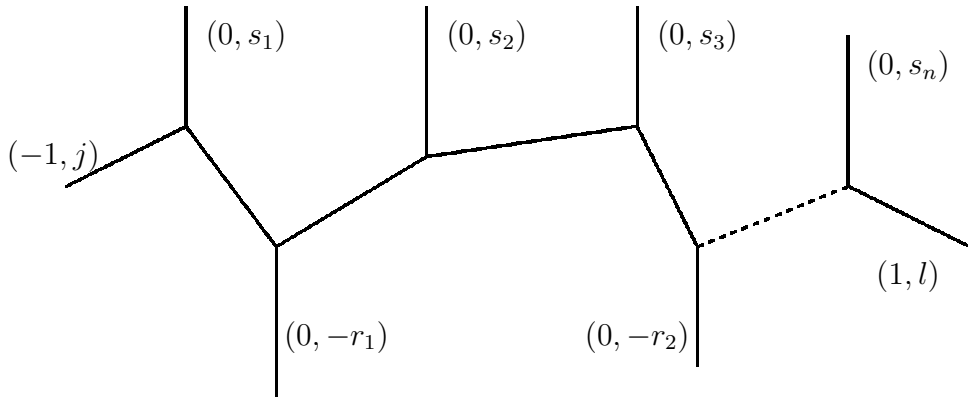
\begin{figure}
\begin{center}
\def\JPicScale{0.8}
\ifx\JPicScale\undefined\def\JPicScale{1}\fi
\unitlength \JPicScale mm
\begin{picture}(160,70)(0,0)
\linethickness{0.3mm}
\multiput(10,40)(0.24,0.12){83}{\line(1,0){0.24}}
\linethickness{0.3mm}
\multiput(30,50)(0.12,-0.16){125}{\line(0,-1){0.16}}
\linethickness{0.3mm}
\multiput(45,30)(0.2,0.12){125}{\line(1,0){0.2}}
\linethickness{0.3mm}
\multiput(70,45)(0.83,0.12){42}{\line(1,0){0.83}}
\linethickness{0.3mm}
\multiput(105,50)(0.12,-0.24){83}{\line(0,-1){0.24}}
\linethickness{0.3mm}
\multiput(140,40)(0.24,-0.12){83}{\line(1,0){0.24}}
\linethickness{0.3mm}
\put(45,5){\line(0,1){25}}
\linethickness{0.3mm}
\put(70,45){\line(0,1){25}}
\linethickness{0.3mm}
\put(105,50){\line(0,1){20}}
\linethickness{0.3mm}
\put(115,10){\line(0,1){20}}
\linethickness{0.3mm}
\put(140,40){\line(0,1){25}}
\linethickness{0.3mm}
\multiput(115,30)(1.85,0.74){14}{\multiput(0,0)(0.31,0.12){3}{\line(1,0){0.31}}}
\linethickness{0.3mm}
\put(30,50){\line(0,1){20}}
\put(8,45){\makebox(0,0)[cc]{$(-1,j)$}}

\put(40,65){\makebox(0,0)[cc]{$(0,s_1)$}}

\put(80,65){\makebox(0,0)[cc]{$(0,s_2)$}}

\put(115,65){\makebox(0,0)[cc]{$(0,s_3)$}}

\put(150,60){\makebox(0,0)[cc]{$(0,s_n)$}}

\put(150,25){\makebox(0,0)[cc]{$(1,l)$}}

\put(55,15){\makebox(0,0)[cc]{$(0, -r_1)$}}

\put(105,15){\makebox(0,0)[cc]{$(0, -r_2)$}}

\end{picture}

\end{center}
\caption{The string network configuration of example 1. 
Here $s_i$ and $r_i$
are positive integers. \label{fn1}}
\end{figure}

\begin{figure}

\begin{center}
\def\JPicScale{0.8}
\ifx\JPicScale\undefined\def\JPicScale{1}\fi
\unitlength \JPicScale mm
\begin{picture}(123.54,53.54)(0,0)
\linethickness{0.3mm}
\put(20,50){\circle{7.07}}

\linethickness{0.3mm}
\put(45,50){\circle{7.07}}

\linethickness{0.3mm}
\put(70,50){\circle{7.07}}

\linethickness{0.3mm}
\put(120,50){\circle{7.07}}

\linethickness{0.3mm}
\put(110,30){\circle{7.07}}

\linethickness{0.3mm}
\put(30,30){\circle{7.07}}

\linethickness{0.3mm}
\put(20,50){\line(1,0){25}}
\linethickness{0.3mm}
\put(45,50){\line(1,0){25}}
\linethickness{0.3mm}
\multiput(110,30)(0.12,0.24){83}{\line(0,1){0.24}}
\linethickness{0.3mm}
\multiput(85,50)(2,0){8}{\line(1,0){1}}
\linethickness{0.3mm}
\multiput(20,50)(0.12,-0.24){83}{\line(0,-1){0.24}}
\linethickness{0.3mm}
\put(75,30){\circle{7.07}}

\linethickness{0.3mm}
\put(30,30){\line(1,0){45}}
\linethickness{0.3mm}
\multiput(30,30)(0.12,0.16){125}{\line(0,1){0.16}}
\linethickness{0.3mm}
\multiput(45,50)(0.18,-0.12){167}{\line(1,0){0.18}}
\linethickness{0.3mm}
\multiput(70,50)(0.12,-0.48){42}{\line(0,-1){0.48}}
\linethickness{0.3mm}
\multiput(90,30)(1.9,0){11}{\line(1,0){0.95}}
\linethickness{0.3mm}
\put(100,50){\circle{7.07}}

\linethickness{0.3mm}
\put(100,50){\line(1,0){20}}
\linethickness{0.3mm}
\multiput(100,50)(0.12,-0.24){83}{\line(0,-1){0.24}}
\linethickness{0.3mm}
\put(85,50){\circle{7.07}}

\linethickness{0.3mm}
\put(90,30){\circle{7.07}}

\linethickness{0.3mm}
\put(70,50){\line(1,0){15}}
\linethickness{0.3mm}
\put(75,30){\line(1,0){15}}
\linethickness{0.3mm}
\multiput(75,30)(0.12,0.24){83}{\line(0,1){0.24}}
\linethickness{0.3mm}
\multiput(85,50)(0.12,-0.48){42}{\line(0,-1){0.48}}
\end{picture}

\end{center}

\vskip -48pt

\caption{The dual grid diagram corresponding to
the network of Fig.\,\ref{fn1}. \label{fn2}}
\end{figure}
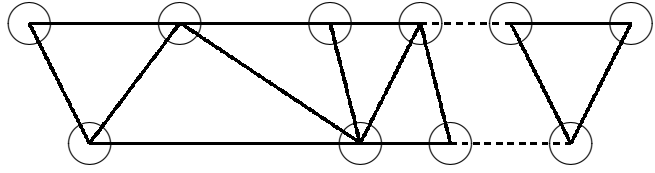

We begin with the planar string network shown in
Fig.\,\ref{fn1}. 
This system has been analyzed extensively in \cite{0005275,0609055}
as bound states of distinct monopoles, 
and we
shall compare our results with the known results later.
The corresponding grid diagram is shown in
Fig.\,\ref{fn2}. The important point to note is that the two
horizontal lines of the grid diagram are separated by unit distance
along the vertical direction and as a result there are no integral
lattice points in the interior of the diagram. Thus we cannot deform
the original network by growing an internal face, and Fig.\,\ref{fn1} represents the
most general network with these external strings.

To compute the index of the state let us approach the wall of marginal
stability where the $(0, s_1)$  string shrinks to zero size. Since the
configuration ceases to exist on the other side of the wall the index
vanishes there and hence the jump in the index across this wall gives the
index of the configuration. This jump on the other hand can be
computed using the primitive wall crossing formula
\refb{eprim}, and the difference in the index of the initial
configuration and that  of the final configuration is given by
 $(-1)^{s_1+1} \, s_1$ (for $B_2(z)$) 
or $(-1)^{s_1+1} \, (y^{s_1} - y^{-s_1})/(y - y^{-1})$
(for $B_1(y,z)$) times the product of the index of a
half BPS state and the index of a quarter BPS network in which the
$(-1,j)$ and $(0,s_1)$ strings are removed. The index of the half BPS
state can be computed from \refb{eindexhalfbps}. On the
other hand to compute the
index of the quarter BPS state we repeat the analysis, this time
approaching the marginal stability wall along which the $(0,-r_1)$
string shrinks to zero size. By repeating this process we can arrive at
the following final expressions for the indices:
\ben \label{efinindex}
B_2(z) &=& (-1)^{\sum s_i +\sum r_j + N-2} 
(z+ z^{-1} - 2)^{N-1} \, \prod_i s_i \prod_j r_j \, , \nn
B_1(y,z) &=& (-1)^{\sum s_i +\sum r_j + N-2} 
\left\{z+z^{-1}-y-y^{-1}\right\}^{N-1} \, \prod_i {y^{s_i}
- y^{-s_i}\over y - y^{-1}} \prod_j {y^{r_j} - y^{-r_j}\over y - y^{-1}} \, , \nn
\een
where $N$ denotes the total number of external strings.

The system described by the string network shown in Fig.\,\ref{fn1} in
fact represents a system of $(N-1)$ distinct monopoles and the supersymmetric
quantum mechanics associated with this system has been thoroughly
analyzed in \cite{0005275,0609055}. In particular Stern and Yi\cite{0005275} 
computed an index in this
supersymmetric quantum mechanics which 
led to a net protected degeneracy of $16\times \prod_i 4s_i \prod_j 4 r_j$.
The value of $B_2(z=-1)$ computed from \refb{efinindex}
is $4\times (-1)^{\sum s_i +\sum r_j +1} \prod_i (4\, s_i) \prod_j 
(4\, r_j)$.
Multiplying the magnitude of this by 4 -- the degeneray due to the
$SU(2)_L$ invariant fermion zero modes
which was factored out from the definition of $B_2(z)$ -- we get the
same result $16\times \prod_i 4s_i \prod_j 4 r_j$.
In fact this result was already rederived in
\cite{0802.0761} by making repeated use of wall crossing formula in 
the manner
we have described above.

There is a more detailed result on Stern-Yi dyon chain in the context of
$\NN=2$ supersymmetric theories due to
Denef\cite{0206072}. 
To compare our result with that of \cite{0206072}, we need to first extract
the result for the Stern-Yi dyon chain in $\NN=2$ supersymmetric theories
from our results.
The dynamics of distinct monopoles in $\NN=2$ supersymmetric theories
can be obtained from those in the $\NN=4$ supersymmetric theories
by projecting out the $SU(2)_L$ non-invariant fermion zero modes
from each constituent monopole. Since each constituent monopole is half-BPS, we
see from table \ref{t1} that the $SU(2)_L$ non-invariant fermion zero modes
transform in the $(2,1,2)$ representation of $SU(2)_L\times SU(2)_R\times SU(2)_{\rm rotation}$
and hence gives a factor of $(z+z^{-1}-y - y^{-1})$ to $B_1$. Since there are 
$(N-1)$ {\it distinct}
constituents these zero modes give a net factor of $(z+z^{-1}-y-y^{-1})^{N-1}$. Factoring out this
contribution from the expression for $B_1$ given in \refb{efinindex} we can get the
result for $B_1(y)$ for the Stern-Yi dyon chain in the $\NN=2$ supersymmetric theory:
\be \label{b1n=2}
B_1(y)|_{\NN=2} = (-1)^{\sum s_i +\sum r_j + N-2} 
\, \prod_i {y^{s_i}
- y^{-s_i}\over y - y^{-1}} \prod_j {y^{r_j} - y^{-r_j}\over y - y^{-1}}\, .
\ee
This agrees with the result of \cite{0206072}.

\begin{figure}
\begin{center}

\def\JPicScale{0.8}
\ifx\JPicScale\undefined\def\JPicScale{1}\fi
\unitlength \JPicScale mm
\begin{picture}(160,70)(0,0)
\linethickness{0.3mm}
\multiput(5,40)(0.3,0.12){83}{\line(1,0){0.3}}
\linethickness{0.3mm}
\multiput(30,50)(0.12,-0.16){125}{\line(0,-1){0.16}}
\linethickness{0.3mm}
\multiput(45,30)(0.2,0.12){125}{\line(1,0){0.2}}
\linethickness{0.3mm}
\multiput(70,45)(0.83,0.12){42}{\line(1,0){0.83}}
\linethickness{0.3mm}
\multiput(105,50)(0.12,-0.24){83}{\line(0,-1){0.24}}
\linethickness{0.3mm}
\multiput(140,40)(0.24,-0.12){83}{\line(1,0){0.24}}
\linethickness{0.3mm}
\put(45,5){\line(0,1){25}}
\linethickness{0.3mm}
\put(70,45){\line(0,1){25}}
\linethickness{0.3mm}
\put(105,50){\line(0,1){20}}
\linethickness{0.3mm}
\put(115,10){\line(0,1){20}}
\linethickness{0.3mm}
\put(140,40){\line(0,1){25}}
\linethickness{0.3mm}
\multiput(115,30)(1.85,0.74){14}{\multiput(0,0)(0.31,0.12){3}{\line(1,0){0.31}}}
\linethickness{0.3mm}
\put(30,50){\line(0,1){20}}
\put(80,65){\makebox(0,0)[cc]{$(0,s_2)$}}

\put(115,65){\makebox(0,0)[cc]{$(0,s_3)$}}

\put(150,60){\makebox(0,0)[cc]{$(0,s_n)$}}

\put(150,25){\makebox(0,0)[cc]{$(1,l)$}}

\put(55,15){\makebox(0,0)[cc]{$(0, -r_1)$}}

\put(105,15){\makebox(0,0)[cc]{$(0, -r_2)$}}

\put(40,65){\makebox(0,0)[cc]{$(0,\ell)$}}

\put(7,35){\makebox(0,0)[cc]{$(-1,j+s_1-\ell)$}}

\end{picture}

\end{center}

\vskip -32pt

\caption{A string network carrying charge vector
$\gamma_1+\ell\gamma_2$ 
that could contribute to the semi-primitive
decay of the network in Fig.\,\ref{fn1}. \label{fn5}}
\end{figure}
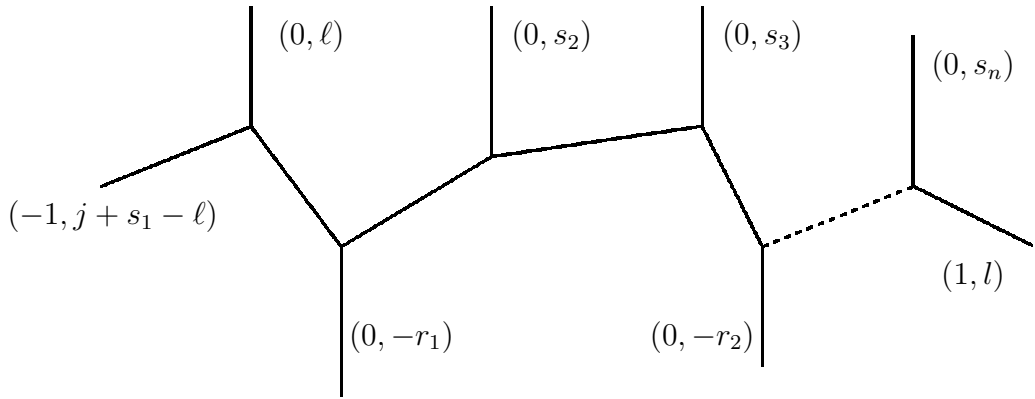

Before leaving this example we note that instead of shrinking the
$(0,s_1)$ string in the first step we could have also shrunk the
$(-1,j)$ string. The wall crossing we shall now encounter is semi-primitive
involving decay into $\gamma_1$ and $s_1\gamma_2$,
with $\gamma_1$ corresponding to the quarter BPS state represented by the
part of the network without the $(-1,j)$ and $(0,s_1)$ string, and
$\gamma_2$ corresponding to the half BPS state represented by
the $(0,1)$ string. Thus $B_2^+(k\gamma_2,z)$ and $B_1^+(k\gamma_2,y,z)$
both vanish for $k>1$ and is given by \refb{eindexhalfbps} for $k=1$.
Furthermore one can check that here $\langle\gamma_1,\gamma_2\rangle=1$.
Thus using \refb{ehalo3}, \refb{ehalo3forb1} we get
\ben\label{eomegahalo}
\sum_{N=0}^\infty \Omega_{\rm halo}(\gamma_1, \gamma_2, N;z) \, q^N
&=& (1+qz) (1+q z^{-1})  (1+q)^{-2}\, ,
\nn
\sum_{N=0}^\infty \Omega_{\rm halo}(\gamma_1, \gamma_2, N;y,z) \, q^N
&=& (1+qz) (1+q z^{-1})  (1+q y)^{-1} (1 + qy^{-1})^{-1}\, .
\een
{}From this we get
\ben \label{eget1}
\Omega_{\rm halo}(\gamma_1, \gamma_2, N;z) &=& (-1)^{N+1} \,
N\, (z+z^{-1}-2)\, ,
\nn
\Omega_{\rm halo}(\gamma_1, \gamma_2, N;y,z) &=& (-1)^{N+1} 
\, \{z+z^{-1}-y-y^{-1}\} \, {y^N - y^{-N}\over y - y^{-1}}\, .
\een
We can now use the semi-primitive wall crossing formul\ae\ \refb{esemib2}
and \refb{esemib1a} with $N=s_1$, 
together with the fact that only $\ell=0$ terms on 
the right hand of these formul\ae\
contribute. To see the latter we have shown in Fig.\,\ref{fn5}
the string network corresponding to the charge vector 
$\gamma_1+\ell\gamma_2$ for $\ell>0$. This has a marginal stability wall
corresponding to shrinking of the $(-1,j+s_1-\ell)$ string and 
in the moduli space this wall coincides with the corresponding wall of
$\gamma_1+s_1\gamma_2$ on which the $(-1,j)$ string 
shown in Fig.\,\ref{fn1} shrinks to zero size. Thus
$B_2^+(\gamma_1+\ell\gamma_2;z)$ and $B_1^+(\gamma_1+\ell\gamma_2;
y,z)$, which correspond to the index measured on the other side of this
wall, vanish for $\ell>0$.  
This shows that the jump in the
index across the wall is 
given by the product of \refb{eget1} with $N$ replaced by $s_1$ 
and the index of $\gamma_1$ -- a quarter BPS state in which the
$(-1,j)$ and $(0,s_1)$ strings are removed from Fig.\,\ref{fn1}. 
This gives us back the same result we have
found before. This provides a consistency check
of our approach and the wall crossing formula.

\subsection{Example 2} 

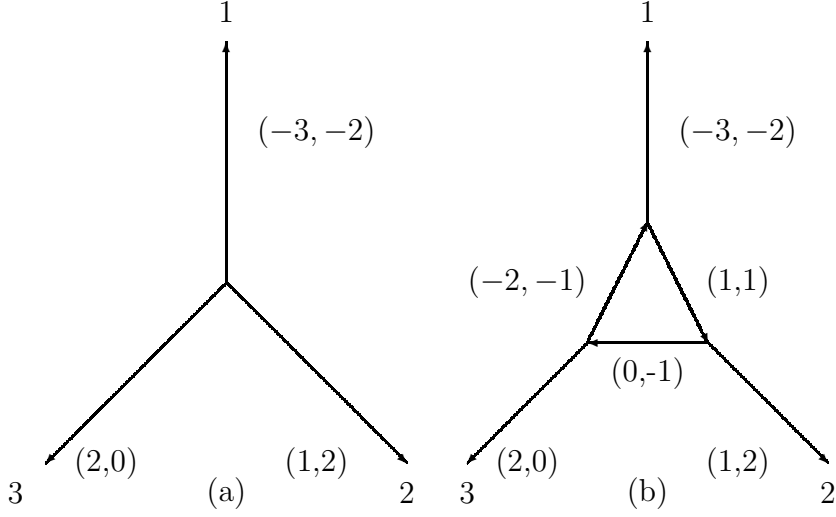
\begin{figure}
\begin{center}
\def\JPicScale{0.8}
\ifx\JPicScale\undefined\def\JPicScale{1}\fi
\unitlength \JPicScale mm
\begin{picture}(140,85)(0,0)
\linethickness{0.3mm}
\put(40,40){\line(0,1){40}}
\put(40,80){\vector(0,1){0.12}}
\linethickness{0.3mm}
\multiput(10,10)(0.12,0.12){250}{\line(1,0){0.12}}
\put(10,10){\vector(-1,-1){0.12}}
\linethickness{0.3mm}
\multiput(40,40)(0.12,-0.12){250}{\line(1,0){0.12}}
\put(70,10){\vector(1,-1){0.12}}
\linethickness{0.3mm}
\multiput(80,10)(0.12,0.12){167}{\line(1,0){0.12}}
\put(80,10){\vector(-1,-1){0.12}}
\linethickness{0.3mm}
\multiput(120,30)(0.12,-0.12){167}{\line(1,0){0.12}}
\put(140,10){\vector(1,-1){0.12}}
\linethickness{0.3mm}
\put(110,50){\line(0,1){30}}
\put(110,80){\vector(0,1){0.12}}
\linethickness{0.3mm}
\multiput(100,30)(0.12,0.24){83}{\line(0,1){0.24}}
\put(110,50){\vector(1,2){0.12}}
\linethickness{0.3mm}
\multiput(110,50)(0.12,-0.24){83}{\line(0,-1){0.24}}
\put(120,30){\vector(1,-2){0.12}}
\linethickness{0.3mm}
\put(100,30){\line(1,0){20}}
\put(100,30){\vector(-1,0){0.12}}
\put(55,65){\makebox(0,0)[cc]{$(-3,-2)$}}

\put(125,65){\makebox(0,0)[cc]{$(-3,-2)$}}

\put(20,10){\makebox(0,0)[cc]{(2,0)}}

\put(55,10){\makebox(0,0)[cc]{(1,2)}}

\put(90,10){\makebox(0,0)[cc]{(2,0)}}

\put(125,10){\makebox(0,0)[cc]{(1,2)}}

\put(90,40){\makebox(0,0)[cc]{$(-2,-1)$}}

\put(125,40){\makebox(0,0)[cc]{(1,1)}}

\put(110,25){\makebox(0,0)[cc]{(0,-1)}}

\put(40,5){\makebox(0,0)[cc]{(a)}}

\put(110,5){\makebox(0,0)[cc]{(b)}}

\put(40,85){\makebox(0,0)[cc]{1}}

\put(110,85){\makebox(0,0)[cc]{1}}

\put(5,5){\makebox(0,0)[cc]{3}}

\put(70,5){\makebox(0,0)[cc]{2}}

\put(80,5){\makebox(0,0)[cc]{3}}

\put(140,5){\makebox(0,0)[cc]{2}}

\end{picture}

\end{center}
\caption{(a) The 
string network configuration of example 2 and (b) its deformation. \label{fn3}}
\end{figure}

We shall now consider the string network shown in Fig.\,\ref{fn3}. The
tree configuration is shown in Fig.\,\ref{fn3}(a), but the analysis of the grid
diagram shows that the network can be deformed to include an internal face
as shown in Fig.\,\ref{fn3}(b). We shall see that including the contribution from
this deformed configuration is essential for the consistency of the wall crossing
formul\ae.

First consider the limit in which the length of the (2,0) string in Fig.\,\ref{fn3}(a)
shrinks to zero size.
In this limit the internal face in Fig.\,\ref{fn3}(b) 
also shrinks to zero size and we reach the
marginal stability wall on which the system becomes unstable against decay into
a pair of half BPS states, one containing a (1,2) string stretched between D3-branes 2 and 3
and a $(-3,-2)$ string stretched between D3-branes 1 and 3. The jump in the index, which also
gives the index since the configuration ceases to exist on the other side of the wall, is
given by the primitive wall crossing formula. The result is
\ben \label{ejj1}
B_2(z) &=& - 4 \, (z + z^{-1} - 2)^2 \, , \nn
B_1(y,z) &=& - \left\{z+z^{-1}-y-y^{-1}\right\}^2 {y^4 - y^{-4}\over y - y^{-1}}\, .
\een

Next consider the limit in which the $(-3,-2)$ string in Fig.\refb{fn3}(a)
shrinks to zero size.
Again in this case neither of the configurations shown in 
Fig.\,\ref{fn3} will survive on the other side of this wall
and hence the index vanishes. Thus the jump in
the index gives the index. However in this case the decay is semi-primitive,
involving the (1,2) string stretched between 1 and 2 and the (2,0) string
stretched between 1 and 3.
Thus we can use the semi-primitive wall crossing formul\ae\ with
$\gamma_1$ representing the (1,2) string stretched between  1 and 2 and
$\gamma_2$ representing the (1,0) string
stretched between 1 and 3, with $\langle\gamma_1,\gamma_2\rangle = 2$.
As in the case of example 1 
$B_2^+(k\gamma_2,z)$ and $B_1^+(k\gamma_2,y,z)$
both vanish for $k>1$ and is given by \refb{eindexhalfbps} for $k=1$.
Thus
using \refb{ehalo3}, \refb{ehalo3forb1} we get
\ben\label{eomegahalonew}
\sum_{N=0}^\infty \Omega_{\rm halo}(\gamma_1, \gamma_2, N;z) \, q^N
&=& (1-qz)^2 (1-q z^{-1})^2  (1-q)^{-4}\, ,
\nn
\sum_{N=0}^\infty \Omega_{\rm halo}(\gamma_1, \gamma_2, N;y,z) \, q^N
&=& (1-qzy) (1-qzy^{-1}) (1-q z^{-1}y) (1-q z^{-1}y^{-1}) \nn &&
(1-q)^{-2} (1-qy^2)^{-1} (1 - qy^{-2})^{-1}\, .
\een
The relevant quantities we need for the decay into $\gamma_1$ and
$2\gamma_2$
are $\Omega_{\rm halo}(\gamma_1, \gamma_2, N=2;z)$
and $\Omega_{\rm halo}(\gamma_1, \gamma_2, N=2;y,z)$.
These can be read out from \refb{eomegahalonew}:
\ben \label{ehaloexp}
\Omega_{\rm halo}(\gamma_1, \gamma_2, N=2;z) &=&
z^2+\frac{1}{z^2}-8 z-\frac{8}{z}+14\, ,
\nn
\Omega_{\rm halo}(\gamma_1, \gamma_2, N=2;y,z) &=&
y^4+\frac{1}{y^4}-y^3 z-\frac{y^3}{z}-\frac{z}{y^3}-\frac{1}{y^3 z}+3
   y^2+\frac{3}{y^2}\nn &&
   -3 y z-\frac{3 y}{z}-\frac{3 z}{y}-\frac{3}{y
   z}+z^2+\frac{1}{z^2}+6\, .
\een
Using  \refb{esemib2}
and \refb{esemib1a} with $N=2$, $\ell=0$ we get
the indices $B_2(z)$ and $B_1(y,z)$, which we shall denote by
$B_2'(z)$ and $B_1'(y,z)$ to distinguish them from \refb{ejj1}.
The results are
\ben \label{ejj2}
B_2'(z) &=& (z+z^{-1}-2) (z^2+\frac{1}{z^2}-8 z-\frac{8}{z}+14)\, , \nn
B_1'(y,z) &=& (z+z^{-1}-y-y^{-1}) \bigg\{y^4+\frac{1}{y^4}-y^3 z
-\frac{y^3}{z}-\frac{z}{y^3}-\frac{1}{y^3 z}+3
   y^2+\frac{3}{y^2}\nn &&
   -3 y z-\frac{3 y}{z}-\frac{3 z}{y}-\frac{3}{y
   z}+z^2+\frac{1}{z^2}+6\bigg\}\, .
\een
These are different from \refb{ejj1}. In fact we have
\ben \label{ejj3}
B_2'(z) - B_2(z) &=& (z+z^{-1}-2)^3 \, , \nn
B_1'(y,z) - B_1(y,z) &=& (z+z^{-1}-y-y^{-1})^3 \, .
\een

\begin{figure}
\begin{center}
\ifx\JPicScale\undefined\def\JPicScale{1}\fi
\unitlength \JPicScale mm
\begin{picture}(80,75)(0,0)
\put(50,75){\makebox(0,0)[cc]{1}}

\put(35,45){\makebox(0,0)[cc]{3}}

\put(80,30){\makebox(0,0)[cc]{2}}

\put(35,60){\makebox(0,0)[cc]{$(-2,-1)$}}

\put(65,60){\makebox(0,0)[cc]{$(1,1)$}}

\put(50,45){\makebox(0,0)[cc]{$(0,-1)$}}

\put(75,40){\makebox(0,0)[cc]{(1,2)}}

\linethickness{0.3mm}
\multiput(40,50)(0.12,0.24){83}{\line(0,1){0.24}}
\put(50,70){\vector(1,2){0.12}}
\linethickness{0.3mm}
\multiput(50,70)(0.12,-0.24){83}{\line(0,-1){0.24}}
\put(60,50){\vector(1,-2){0.12}}
\linethickness{0.3mm}
\put(40,50){\line(1,0){20}}
\put(40,50){\vector(-1,0){0.12}}
\linethickness{0.3mm}
\multiput(60,50)(0.12,-0.16){125}{\line(0,-1){0.16}}
\put(75,30){\vector(3,-4){0.12}}
\end{picture}
\end{center}

\vskip -100pt

\caption{A wall of marginal stability of the string network
of Fig.\,\ref{fn3}(b).
\label{fn4}}
\end{figure}
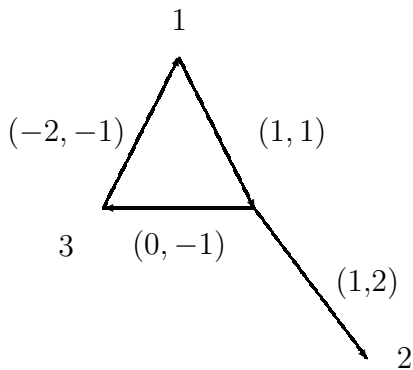

The fact that $B_2(z), B_1(y,z)$ are different from $B_2'(z), B_1'(y,z)$
is not an immediate contradiction since they represent indices computed
in different regions in the moduli space -- the former in a region where
the (2,0) string is short and the latter in a region where the $(-3,-2)$
string is short. However for consistency we need to show that these two
regions are separated by a new wall of marginal stability and that the
jump in the index across this wall accounts for the differences shown in
\refb{ejj3}. This new wall can be identified by considering the network
shown in Fig.\,\ref{fn3}(b). When the $(-3,-2)$ string is short then the
maximal size of the internal face is set by the configuration where the face
touches the D3-brane 1. On the other hand when the (2,0) string is
short then the maximal size of the internal face is set by the 
configuration where the face
touches the D3-brane 3. The boundary between these two regions
of the moduli space corresponds to an arrangement of the D3-branes
1 and 3 such that when the internal face touches the D3-brane 1 it also touches
the D3-brane 3.\footnote{During this deformation of the moduli we can keep
the D3-brane 2 far away so that the internal face never touches it.} This
situation has been shown in Fig\,\ref{fn4}. From this diagram it is clear
that this represents a wall of marginal stability along which the
original network is unstable against decay into a $(-2,-1)$ string
stretched between D3-branes 1 and 3 and the rest of the network
containing the $(1,1)$, $(0,-1)$ and (1,2) strings. The jump across this
wall can be computed using the primitive wall crossing formula
and involves the product of the index of a half BPS state
represented by the $(-2,-1)$ string and a quarter BPS state containing the
$(1,1)$, $(0,-1)$ and (1,2) strings. The former is known from
\refb{eindexhalfbps} while the latter can be found by applying the
wall crossing formula again
across the wall on which the $(0,-1)$ string shrinks to  zero size.
The result is the following expression for the jump in the index
across the marginal stability wall shown in Fig.\,\ref{fn4}:
\ben \label{ejj4}
\Delta B_2(z) &=& (z+z^{-1}-2)^3 \, , \nn
\Delta B_1(y,z) &=& (z+z^{-1}-y-y^{-1})^3 \, .
\een
This accounts for the difference \refb{ejj3}. By carefully calculating the
phase of $Z_\gamma$ one can verify that 
\refb{ejj4} actually represents the jump in the index that we encounter
as we cross from the side in which 
the
(2,0) string is short towards the side on which the $(-3,-2)$ string is short.
This
is precisely what is needed to 
explain the difference between $B_2'$, $B_1'$ and $B_2$, $B_1$ given in
\refb{ejj3}.

\subsection{Example 3}

\begin{figure}
\begin{center}
\def\JPicScale{0.6}
\ifx\JPicScale\undefined\def\JPicScale{1}\fi
\unitlength \JPicScale mm
\begin{picture}(110,80)(0,0)
\linethickness{0.3mm}
\multiput(20,80)(0.16,-0.12){250}{\line(1,0){0.16}}
\linethickness{0.3mm}
\multiput(60,50)(0.19,0.12){208}{\line(1,0){0.19}}
\linethickness{0.3mm}
\put(60,15){\line(0,1){35}}
\put(45,75){\makebox(0,0)[cc]{(2,0)}}

\put(75,30){\makebox(0,0)[cc]{(3,15)}}

\put(110,60){\makebox(0,0)[cc]{$(-5, -15)$}}

\end{picture}

\end{center}

\vskip -48pt

\caption{The
string network configuration of example 3. \label{fks5}}
\end{figure}
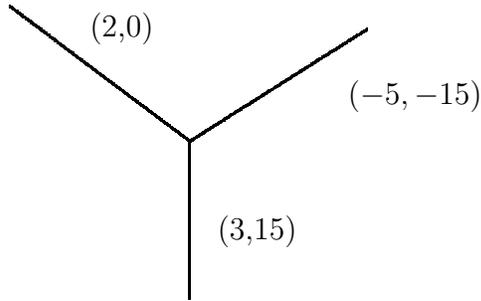

The final example we shall consider is the string network shown in
Fig.\,\ref{fks5} and its possible deformations. We shall compute the
index in the chamber in which the $(-5,-15)$ string is short, \i.e.\ near
the wall where it can break apart into $(2,0)=2(1,0)$ and $(3,15)=3(1,5)$ 
string. 
For brevity we shall only compute the index $B_1(y,z)$ since
$B_2(z)$ can be obtained by taking the $y\to 1$ limit of $B_1(y,z)$.
Since the decay across the wall on which the $(-5,-15)$ string shrinks to
zero size is neither primitive nor semi-primitive, we need the  full
power of the KS wall crossing formula.
Labelling by $\gamma_1$ the charge carried by the $(1,0)$ string and
by $\gamma_2$ the charge carried by the $(1,5)$ string, we see that
\be \label{egamma}
\gamma \equiv \langle \gamma_1, \gamma_2\rangle = -5\, .
\ee
We also define
\be \label{ekapgam}
\kappa(x) = (-1)^x {y^x - y^{-x}\over y - y^{-1}}\, .
\ee
Now the index we want to compute is $B_1^-(2\gamma_1+3\gamma_2;y, z)$.
Using logic similar to the one used in the earlier examples
we see that on the other side (+ side) of the wall of marginal stability
the only non-zero indices of relevance are those of the half BPS states
carrying charges $\gamma_1$ or $\gamma_2$:
\ben \label{enonzero}
&& B_1^+(\gamma_1; y,z) = B_1^+(\gamma_2;y,z) = z + z^{-1} - y - y^{-1}\, ,
\nn &&
B_1^+(m\gamma_1+n\gamma_2)=0 \quad \hbox{otherwise}\, .
\een
Eq.\refb{emot1} now gives
\ben \label{eratnon}
&& \bar B_1^+(m\gamma_1; y,z) = \bar B_1^+(m\gamma_2;y,z) =
\left(z^m + z^{-m} - y^m - y^{-m}\right)\, {1\over m} \, {y - y^{-1}\over y^m - y^{-m}}
\, , \nn
&& \bar B_1^+(m\gamma_1+n\gamma_2)=0 \quad \hbox{otherwise}\, .
\een
Using eq.(A.4) of \cite{1011.1258} we now get
\ben \label{efinexp}
B_1^-(2\gamma_1+3\gamma_2; y,z) &=& 
\kappa(6\gamma) \bar B_1^+(2\gamma_1; y,z) \bar B_1^+(3\gamma_2; y,z)
+ {1\over 2} \kappa(3\gamma)^2 \bar B_1^+(\gamma_1; y,z)^2 
\bar B_1^+(3\gamma_2; y,z)
\nn &&
+ \kappa(2\gamma) \kappa(4\gamma) 
\bar B_1^+(2\gamma_1; y,z) \bar B_1^+(2\gamma_2; y,z)\bar 
B_1^+(\gamma_2; y,z)\nn
&& + {1\over 2} \kappa(\gamma) \kappa(2\gamma) \{\kappa(\gamma)
+ \kappa(3\gamma)\} \bar B_1^+(\gamma_1; y,z)^2 \bar B_1^+(2\gamma_2; y,z)
\bar B_1^+(\gamma_2; y,z) \nn &&
+{1\over 6} \kappa(2\gamma)^3 \bar B_1^+(2\gamma_1; y,z)
\bar B_1^+(\gamma_2; y,z)^3 \nn &&
+{1\over 12} \kappa(\gamma)^3 \{3\kappa(\gamma) +\kappa(3\gamma)\}
\bar B_1^+(\gamma_1; y,z)^2 \bar B_1^+(\gamma_2; y,z)^3\, . 
\een
Eqs.\refb{egamma}-\refb{efinexp} gives us the complete expression
for $B_1^-(2\gamma_1+3\gamma_2;y,z)$.

\bigskip

{\bf Acknowledgement:} 
I would like to thank Jan Manschot and Boris Pioline for 
useful comments on an earlier
version of the manuscript.
This work was
supported in part by the project 11-R\&D-HRI-5.02-0304
and the J. C. Bose fellowship of 
the Department of Science and Technology, India.


\small

\baselineskip 12pt


\begin{thebibliography}{99}

\bibitem{wittenolive} 
  E.~Witten and D.~I.~Olive,
  ``Supersymmetry Algebras That Include Topological Charges,''
  Phys.\ Lett.\ B {\bf 78}, 97 (1978).
 
 \bibitem{osborn} 
  H.~Osborn,
  ``Topological Charges for N=4 Supersymmetric Gauge Theories and Monopoles of Spin 1,''
  Phys.\ Lett.\ B {\bf 83}, 321 (1979).
  
 \bibitem{9402032} 
  A.~Sen,
  ``Dyon - monopole bound states, selfdual harmonic forms on the multi - monopole moduli space, and SL(2,Z) invariance in string theory,''
  Phys.\ Lett.\ B {\bf 329}, 217 (1994)
  [hep-th/9402032].

\bibitem{segalselby}
G.~Segal and A.~Selby,
``The cohomology of the space of magnetic monopoles,''
Comm.\ Math.\ Phys.\ {\bf 177}, 775 (1996).

\bibitem{9712211}
  O.~Bergman,
   ``Three-pronged strings and 1/4 BPS states in N=4 super-Yang-Mills
  theory,''
  Nucl.\ Phys.\ B {\bf 525}, 104 (1998)
  [arXiv:hep-th/9712211].

\bibitem{9804160}
  O.~Bergman and B.~Kol,
  ``String webs and 1/4 BPS monopoles,''
  Nucl.\ Phys.\  B {\bf 536}, 149 (1998)
  [arXiv:hep-th/9804160].

\bibitem{9607201}
  J.~H.~Schwarz,
  ``Lectures on superstring and M theory dualities: Given at ICTP Spring School and at TASI Summer School,''
  Nucl.\ Phys.\ Proc.\ Suppl.\  {\bf 55B} (1997) 1
  [hep-th/9607201].

\bibitem{9704170} 
  O.~Aharony and A.~Hanany,
  ``Branes, superpotentials and superconformal fixed points,''
  Nucl.\ Phys.\ B {\bf 504}, 239 (1997)
  [hep-th/9704170].

\bibitem{9710116}
  O.~Aharony, A.~Hanany and B.~Kol,
  ``Webs of (p,q) five-branes, five-dimensional field theories and grid diagrams,''
  JHEP {\bf 9801} (1998) 002
  [hep-th/9710116].

\bibitem{9711094}
  K.~Dasgupta and S.~Mukhi,
  ``BPS nature of three string junctions,''
  Phys.\ Lett.\ B {\bf 423} (1998) 261
  [hep-th/9711094].

\bibitem{9711130}
  A.~Sen,
  ``String network,''
  JHEP {\bf 9803} (1998) 005
  [hep-th/9711130].
  
 \bibitem{9812021} 
  B.~Kol,
  ``Thermal monopoles,''
  JHEP {\bf 0007}, 026 (2000)
  [hep-th/9812021].

\bibitem{9804174}
  K.~M.~Lee and P.~Yi,
  ``Dyons in N = 4 supersymmetric theories and three-pronged strings,''
  Phys.\ Rev.\  D {\bf 58}, 066005 (1998)
  [arXiv:hep-th/9804174].


\bibitem{9907090}
  D.~Bak, K.~M.~Lee and P.~Yi,
  ``Quantum 1/4 BPS dyons,''
  Phys.\ Rev.\  D {\bf 61}, 045003 (2000)
  [arXiv:hep-th/9907090].


\bibitem{9912082} 
  J.~P.~Gauntlett, N.~Kim, J.~Park and P.~Yi,
  ``Monopole dynamics and BPS dyons N=2 superYang-Mills theories,''
  Phys.\ Rev.\ D {\bf 61}, 125012 (2000)
  [hep-th/9912082].

\bibitem{0005275}
  M.~Stern and P.~Yi,
  ``Counting Yang-Mills dyons with index theorems,''
  Phys.\ Rev.\  D {\bf 62}, 125006 (2000)
  [arXiv:hep-th/0005275].


\bibitem{0609055}
  E.~J.~Weinberg and P.~Yi,
  ``Magnetic monopole dynamics, supersymmetry, and duality,''
  Phys.\ Rept.\  {\bf 438}, 65 (2007)
  [arXiv:hep-th/0609055].
  
\bibitem{0712.3625}
  K.~Narayan,
  ``On the internal structure of dyons in ${\cal N}=4$ super Yang-Mills
  theories,''
  arXiv:0712.3625 [hep-th].

\bibitem{0802.0761}
A.~Dabholkar, K.~Narayan and S.~Nampuri, 
``Degeneracy of Decadent Dyons,''
arXiv:0802.0761[hep-th].
  
\bibitem{9211097}
  S.~Cecotti, C.~Vafa,
  ``On classification of N=2 supersymmetric theories,''
  Commun.\ Math.\ Phys.\  {\bf 158}, 569-644 (1993).
  [hep-th/9211097].

\bibitem{9407087}
  N.~Seiberg, E.~Witten,
  ``Electric - magnetic duality, monopole condensation, 
  and confinement in N=2 supersymmetric Yang-Mills theory,''
  Nucl.\ Phys.\  {\bf B426}, 19-52 (1994).
  [hep-th/9407087].

\bibitem{9408099}
  N.~Seiberg, E.~Witten,
  ``Monopoles, duality and chiral symmetry breaking 
  in N=2 supersymmetric QCD,''
  Nucl.\ Phys.\  {\bf B431}, 484-550 (1994).
  [hep-th/9408099].

\bibitem{9602082}
  F.~Ferrari, A.~Bilal,
  ``The Strong coupling spectrum of the Seiberg-Witten theory,''
  Nucl.\ Phys.\  {\bf B469 } (1996)  387-402.
  [hep-th/9602082].
  
 \bibitem{9605101}
  A.~Bilal, F.~Ferrari,
  ``Curves of marginal stability, and weak and strong 
  coupling BPS spectra in N=2 supersymmetric QCD,''
  Nucl.\ Phys.\  {\bf B480}, 589-622 (1996).
  [hep-th/9605101].

\bibitem{9902116}
  E.~R.~Sharpe,
  ``D-branes, derived categories, and Grothendieck groups,''
  Nucl.\ Phys.\ B {\bf 561} (1999) 433
  [hep-th/9902116].

\bibitem{0011017}
  M.~R.~Douglas,
  ``D-branes, categories and N=1 supersymmetry,''
  J.\ Math.\ Phys.\  {\bf 42}, 2818-2843 (2001).
  [hep-th/0011017].


\bibitem{0005049}
  F.~Denef,
  ``Supergravity flows and D-brane stability,''
  JHEP {\bf 0008}, 050 (2000).
  [hep-th/0005049].

\bibitem{0206072}
  F.~Denef,
  ``Quantum quivers and Hall / hole halos,''
  JHEP {\bf 0210}, 023 (2002).
  [arXiv:hep-th/0206072 [hep-th]].

\bibitem{0304094}
  B.~Bates, F.~Denef,
  ``Exact solutions for supersymmetric stationary black hole composites,''
  [hep-th/0304094].

\bibitem{0702141}
  A.~Sen,
  ``Walls of Marginal Stability and Dyon Spectrum in N=4 Supersymmetric String Theories,''
  JHEP {\bf 0705} (2007) 039
  [hep-th/0702141].

\bibitem{0702146}
  F.~Denef, G.~W.~Moore,
  ``Split states, entropy enigmas, holes and halos,''
  [arXiv:hep-th/0702146 [HEP-TH]].

\bibitem{0702150}
  A.~Dabholkar, D.~Gaiotto and S.~Nampuri,
  ``Comments on the spectrum of CHL dyons,''
  JHEP {\bf 0801} (2008) 023
  [hep-th/0702150 [HEP-TH]].

\bibitem{0705.3874}
  A.~Sen,
  ``Two centered black holes and N=4 dyon spectrum,''
  JHEP {\bf 0709} (2007) 045
  [arXiv:0705.3874 [hep-th]].

\bibitem{0706.2363}
  M.~C.~N.~Cheng and E.~Verlinde,
  ``Dying Dyons Don't Count,''
  JHEP {\bf 0709} (2007) 070
  [arXiv:0706.2363 [hep-th]].

\bibitem{0811.2435}
M.~Kontsevich and Y.~Soibelman, ``{Stability structures, motivic
  Donaldson-Thomas invariants and cluster transformations},''
 [arXiv:0811.2435 [math.AG]]. 
 
\bibitem{0910.4315}
  M.~Kontsevich, Y.~Soibelman,
  ``Motivic Donaldson-Thomas invariants: Summary of results,''
  [arXiv:0910.4315 [math.AG]].

\bibitem{1006.2706}
  M.~Kontsevich, Y.~Soibelman,
  ``Cohomological Hall algebra, exponential Hodge structures and 
  motivic Donaldson-Thomas invariants,''
  [arXiv:1006.2706 [math.AG]].

\bibitem{0410268}
D.~Joyce, ``Configurations in abelian categories. {IV}. {I}nvariants and
  changing stability conditions,'' {\em Adv. Math.} {\bf 217} (2008), no.~1,
  125--204 [ arXiv:math/0410268].

\bibitem{0810.5645}
  D.~Joyce, Y.~Song,
 ``A Theory of generalized Donaldson-Thomas invariants,''
  [arXiv:0810.5645 [math.AG]].

\bibitem{0910.0105}
  D.~Joyce,
  ``Generalized Donaldson-Thomas invariants,''
  [arXiv:0910.0105 [math.AG]].

\bibitem{0706.3193}
  E.~Diaconescu, G.~W.~Moore,
  ``Crossing the wall: Branes versus bundles,''
  [arXiv:0706.3193 [hep-th]].

\bibitem{0807.4723}
  D.~Gaiotto, G.~W.~Moore, A.~Neitzke,
  ``Four-dimensional wall-crossing via three-dimensional field theory,''
  Commun.\ Math.\ Phys.\  {\bf 299}, 163-224 (2010).
  [arXiv:0807.4723 [hep-th]].

\bibitem{0812.4219}
  S.~Alexandrov, B.~Pioline, F.~Saueressig, S.~Vandoren,
  ``D-instantons and twistors,''
  JHEP {\bf 0903}, 044 (2009).
  [arXiv:0812.4219 [hep-th]].

\bibitem{0807.4556}
J.~de~Boer, S.~El-Showk, I.~Messamah, and D.~Van~den Bleeken, ``{Quantizing N=2
  Multicenter Solutions},'' {\em JHEP} {\bf 05} (2009) 002,
\href{http://www.arXiv.org/abs/0807.4556}{{\tt 0807.4556}}.

\bibitem{0810.4909}
  D.~L.~Jafferis, G.~W.~Moore,
  ``Wall crossing in local Calabi Yau manifolds,'' 
  [arXiv:0810.4909 [hep-th]].

\bibitem{0904.1420}
T.~Dimofte and S.~Gukov, ``{Refined, Motivic, and Quantum},'' {\em Lett. Math.
  Phys.} {\bf 91} (2010) 1,
\href{http://www.arXiv.org/abs/0904.1420}{{\tt 0904.1420}}.

\bibitem{0910.2615}
  S.~Cecotti, C.~Vafa,
  ``BPS Wall Crossing and Topological Strings,''  
  [arXiv:0910.2615 [hep-th]].

\bibitem{0912.2923}
J.~Stoppa, ``{D0-D6 states counting and GW invariants},''
\href{http://www.arXiv.org/abs/0912.2923}{{\tt 0912.2923}}.
   

\bibitem{0912.2507}
Y.~Toda, ``{On a computation of rank two Donaldson-Thomas invariants},''
\href{http://www.arXiv.org/abs/0912.2507}{{\tt 0912.2507}}.

\bibitem{1002.0579}
W.-y. Chuang, D.-E. Diaconescu, and G.~Pan, ``{Rank Two ADHM Invariants and
  Wallcrossing},''
\href{http://www.arXiv.org/abs/1002.0579}{{\tt 1002.0579}}.

\bibitem{1006.3435}
  S.~Cecotti, A.~Neitzke, C.~Vafa,
  ``R-Twisting and 4d/2d Correspondences,'
  [arXiv:1006.3435 [hep-th]].

\bibitem{1006.0146}
  D.~Gaiotto, G.~W.~Moore, A.~Neitzke,
  ``Framed BPS States,''  
  [arXiv:1006.0146 [hep-th]].

\bibitem{1008.0030}
  E.~Andriyash, F.~Denef, D.~L.~Jafferis, G.~W.~Moore,
  ``Wall-crossing from supersymmetric galaxies,''  
  [arXiv:1008.0030 [hep-th]].

\bibitem{0906.1767}
  J.~Manschot,
  ``Stability and duality in N=2 supergravity,''
  Commun.\ Math.\ Phys.\  {\bf 299}, 651-676 (2010).
  [arXiv:0906.1767 [hep-th]].

\bibitem{0912.1346}
T.~Dimofte, S.~Gukov, and Y.~Soibelman, ``{Quantum Wall Crossing in N=2 Gauge
  Theories},'' {\em Lett. Math. Phys.} {\bf 95} (2011) 1-25 
\href{http://www.arXiv.org/abs/0912.1346}{{\tt 0912.1346}}.

\bibitem{1011.1258}
  J.~Manschot, B.~Pioline, A.~Sen,
  ``Wall Crossing from Boltzmann Black Hole Halos,''
  JHEP {\bf 1107}, 059 (2011).
  [arXiv:1011.1258 [hep-th]].

\bibitem{1003.1570}
J.~Manschot, ``{Wall-crossing of D4-branes using flow trees},''
\href{http://www.arXiv.org/abs/1003.1570}{{\tt 1003.1570}}.

\bibitem{1009.1775}
J.~Manschot, ``{The Betti numbers of the moduli space of stable sheaves of rank
  3 on P2},''
\href{http://www.arXiv.org/abs/1009.1775}{{\tt 1009.1775}}.

\bibitem{1010.6002}
  T.~Nishinaka,
  ``Multiple D4-D2-D0 on the Conifold and Wall-crossing with the Flop,''
  \href{http://www.arXiv.org/abs/1010.6002}{{\tt 1010.6002}}.
  
\bibitem{1102.1729} 
  S.~Lee and P.~Yi,
  ``Framed BPS States, Moduli Dynamics, and Wall-Crossing,''
  JHEP\ {\bf 1104}, 098  (2011)
  [arXiv:1102.1729 [hep-th]].

\bibitem{1103.0261}
  B.~Pioline,
 ``Four ways across the wall,''
  [arXiv:1103.0261 [hep-th]].
  
  \bibitem{1103.1887}
  J.~Manschot, B.~Pioline, A.~Sen,
  ``A Fixed point formula for the index of multi-centered N=2 black holes,''
  JHEP {\bf 1105}, 057 (2011).
  [arXiv:1103.1887 [hep-th]].

\bibitem{1107.0723}
  H.~Kim, J.~Park, Z.~Wang, P.~Yi,
  ``Ab Initio Wall-Crossing,''
  JHEP {\bf 1109}, 079 (2011).
  [arXiv:1107.0723 [hep-th]].

\bibitem{1109.4861} 
  J.~Manschot,
  ``BPS invariants of semi-stable sheaves on $p^2$ and its blow-up,''
  arXiv:1109.4861 [math-ph].
  
\bibitem{1109.4941} 
  M.~Alim, S.~Cecotti, C.~Cordova, S.~Espahbodi, A.~Rastogi and C.~Vafa,
  ``BPS Quivers and Spectra of Complete N=2 Quantum Field Theories,''
  arXiv:1109.4941 [hep-th].
  
  
  \bibitem{1110.0466}
  S.~Alexandrov, D.~Persson, B.~Pioline,
  ``Wall-crossing, Rogers dilogarithm, and the QK/HK correspondence,''
  [arXiv:1110.0466 [hep-th]].

\bibitem{1110.1619} 
  A.~Neitzke,
  ``On a hyperholomorphic line bundle over the Coulomb branch,''
  arXiv:1110.1619 [hep-th].

\bibitem{1111.6979} 
  D.~V.~d.~Bleeken,
  ``BPS dyons and Hesse flow,''
  arXiv:1111.6979 [hep-th].

\bibitem{1112.2174} 
J.~Stoppa, ``Joyce-Song wall-crossing as an asymptotic expansion,"
 arXiv:1112.2174 [math.AG].
  
\bibitem{1112.2515} 
  A.~Sen,
  ``Equivalence of Three Wall Crossing Formulae,''
  arXiv:1112.2515 [hep-th].

\bibitem{9611205}
  C.~Bachas and E.~Kiritsis,
  ``F(4) terms in N=4 string vacua,''
  Nucl.\ Phys.\ Proc.\ Suppl.\  {\bf 55B} (1997) 194
  [hep-th/9611205].

\bibitem{9708062}
  A.~Gregori, E.~Kiritsis, C.~Kounnas, N.~A.~Obers, P.~M.~Petropoulos and B.~Pioline,
  ``R**2 corrections and nonperturbative dualities of N=4 string ground states,''
  Nucl.\ Phys.\ B {\bf 510} (1998) 423
  [hep-th/9708062].

\bibitem{0802.1556}
  S.~Banerjee, A.~Sen and Y.~K.~Srivastava,
  ``Partition Functions of Torsion > 1 Dyons in Heterotic String Theory on T**6,''
  JHEP {\bf 0805} (2008) 098
  [arXiv:0802.1556 [hep-th]].

\bibitem{0803.3857}
  A.~Sen,
  ``Wall Crossing Formula for N=4 Dyons: A Macroscopic Derivation,''
  JHEP {\bf 0807} (2008) 078
  [arXiv:0803.3857 [hep-th]].

\bibitem{0911.1563}
  A.~Sen,
  ``A Twist in the Dyon Partition Function,''
  JHEP {\bf 1005} (2010) 028
  [arXiv:0911.1563 [hep-th]].

\bibitem{1002.3857}
  A.~Sen,
 ``Discrete Information from CHL Black Holes,''
  JHEP {\bf 1011} (2010) 138
  [arXiv:1002.3857 [hep-th]].

\bibitem{0707.1563}
  A.~Sen,
  ``Rare Decay Modes of Quarter BPS Dyons,''
  JHEP {\bf 0710} (2007) 059
  [arXiv:0707.1563 [hep-th]].

\bibitem{0809.1157}
  S.~Mukhi and R.~Nigam,
  ``Constraints on 'rare' dyon decays,''
  JHEP {\bf 0812} (2008) 056
  [arXiv:0809.1157 [hep-th]].

\bibitem{0903.2481}
  A.~Dabholkar, M.~Guica, S.~Murthy and S.~Nampuri,
  ``No entropy enigmas for N=4 dyons,''
  JHEP {\bf 1006} (2010) 007
  [arXiv:0903.2481 [hep-th]].



\end{thebibliography}
\end{document}